\documentclass[twocolumn,3p]{elsarticle}
\usepackage{amsmath}
\usepackage{amsfonts}
\usepackage{amssymb}
\usepackage{graphicx}

\begin{document}

\title{PAIRING SYMMETRY AND PAIRING STATE IN FERROPNICTIDES: THEORETICAL
OVERVIEW}
\author{I.I. Mazin$^{a}$ and J. Schmalian$^{b}$ }
\date{February 17, 2009}

\begin{abstract}
We review the main ingredients for an unconventional pairing state in the
ferropnictides, with particular emphasis on interband pairing due to
magnetic fluctuations. Summarizing the key experimental prerequisites for
such pairing, the electronic structure and nature of magnetic excitations,
we discuss the properties of the $s^{\pm}$ state that emerges as a likely
candidate pairing state for these materials and survey experimental evidence
in favor of and against this novel state of matter.
\end{abstract}
\address[els]{ Code 6391, Naval Research Laboratory, Washington,
DC 20375} 
\address[els]{ Iowa State University and Ames Laboratory,
Ames, IA, 50011 }
\maketitle

\begin{flushright}
\textit{One fist of iron, the other of steel\newline
If the right one don't get you, then the left one will\newline
Merle Travis, 16 tons}
\end{flushright}

\section{Introduction}

The discovery of cuprate superconductors has changed our mentality in many
ways. In particular, the question that would have sounded moot to most
before 1988, \textit{what is the symmetry of the superconducting state}, is
now the first question to be asked when a new superconductor has been
discovered. The pool of potential candidates, before considered at best a
mental Tetris for theorists, had acquired a practical meaning. It has been
demonstrated that superconductivity in cuprates is $d$-wave, while in MgB$%
_{2}$ it is multi-gap $s$-wave with a large gap disparity. There is
considerable evidence that Sr$_{2}$RuO$_{4}$ is a $p$-wave material. Other
complex order parameters are routinely discussed for heavy fermion systems
or organic charge transfer salts. It is likely that the newly discovered
ferropnictides represent another superconducting state, not encountered in
experiment before.

Besides the general appreciation that pairing states may be rather
nontrivial, it has also been recognized that unconventional pairing is
likely due, at least to some extent, to electronic (Coulomb or magnetic)
mechanisms and, conversely, electronic mechanisms are much more likely to
produce unconventional pairing symmetries than the standard uniform-gap $s$%
-wave. It has been appreciated that the actual symmetry is very sensitive to
the momentum dependence of the pairing interaction, as well as to the
underlying electronic structure (mostly, fermiology).

Therefore we have structured this overview so that it starts with a layout
of prerequisites for a meaningful discussion of the pairing symmetry. First
of all, we shall describe the gross features of the fermiology according to
density-functional (DFT) calculations, as well as briefly assess
verification of such calculations \textit{via} ARPES and quantum
oscillations experiments. Again, detailed discussion of these can be found
elsewhere in this volume. We will also point out where one may expect
caveats in using the DFT band structure: it is in our view misleading to
assume that these compounds are uncorrelated. While not necessarily of the
same nature as in cuprates, considerable electron-electron interaction
effects cannot be excluded and are even expected.

We will then proceed to discuss the role of magnetic fluctuations as well as
other excitations due to electron-electron interactions. We discuss the
special role the antiferromagnetic (AFM) ordering vector plays for the
pairing symmetry and address the on-site Coulomb (Hubbard correlations), to
the extent of their possible effect on the pairing symmetry, and possible
overscreeining (Ginzburg-Little) interactions. We also discuss puzzling
issues that are related to the magnetoelastic interaction in these systems.
As for a discussion of the electron-phonon interaction we refer to the
article by Boeri \textit{et al} in this volume. The final part of this
review consists of a summary of theoretical aspects of the pairing state,
along with a discussion of its experimental manifestations.

\section{Prerequisites for addressing the Cooper pairing}

\subsection{Electronic structure and fermiology}

\subsubsection{Density functional calculations}

\begin{figure}[ptb]
\includegraphics[width=0.9\linewidth,viewport=40 100 460
450,clip]{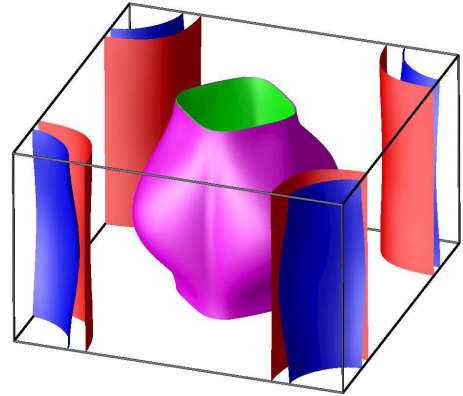}
\caption{(color online) The Fermi surface of undoped nonmagnetic FeTe. 
\protect\cite{Ref1}}
\label{FeTe}
\end{figure}

The two families of the Fe-based superconductors are $1111$ systems ROFeAs
with rare earth ions R\cite{Jeitschko,Kamihara} and the $122$ systems AFe$%
_{2}$As$_{2}$ with alkaline earth element A\cite{Rotter08}. Both families
have been studied in much detail by first principles DFT calculations. Here
and below, unless specifically indicated, we use a 2D unit cell with two Fe
per cell, and the corresponding reciprocal lattice cell; the $x$ and $y$
directions are along the next-nearest-neighbor Fe-Fe bond. It appears that
all materials share the same common motif: two or more hole-like Fermi
surfaces near the $\Gamma $ point [$\mathbf{k=}(0,0)$], and two
electron-like surfaces near the M point [$\mathbf{k=}(\pi ,\pi )$] (Fig. \ref%
{FS1}-\ref{FeTe}). This is true, however, in strictly non-magnetic
calculations only, when the magnetic moment on each Fe is restricted to
zero. As discussed below, this is not necessarily a correct picture.

\begin{figure}[tbp]
\includegraphics[width=0.90\linewidth]{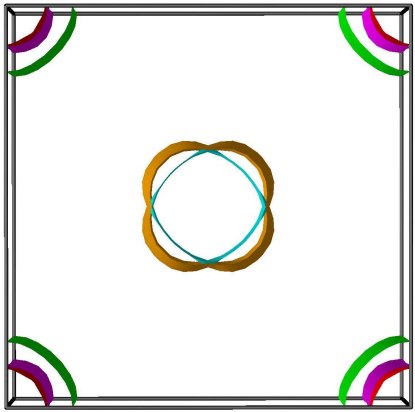}
\caption{(color online) The Fermi surface of the non-magnetic LaAsFeO for
10\% e-doping \protect\cite{Ref1}}
\label{FS1}
\end{figure}
\begin{figure}[tbp]
\includegraphics[width=0.99\linewidth,viewport=-20 20 500 480,clip
]{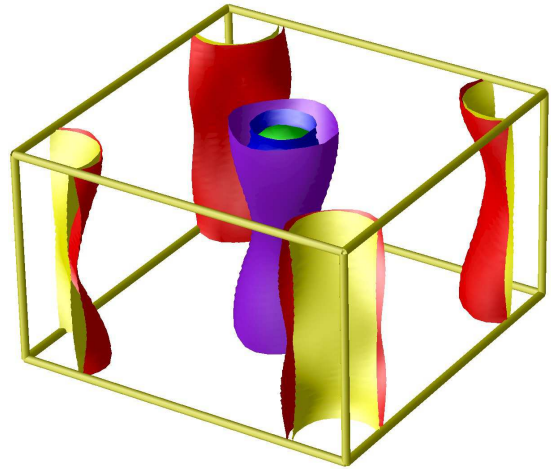}
\caption{(color online) The Fermi surface of the non-magnetic BaFe$_{2}$As$%
_{2}$ for 10\% e-doping (Co doping, virtual crysatl approximation)\protect\cite{Ref1}}
\label{FS2}
\end{figure}

\begin{figure}[ptb]
\includegraphics[width=0.9\linewidth,viewport=40 150 460
410,clip]{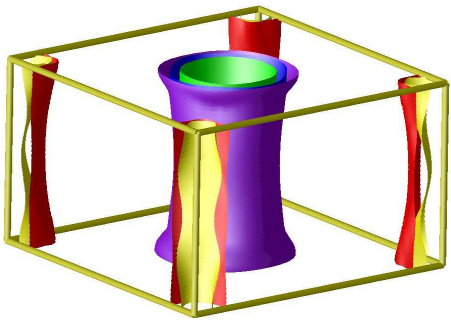}
\caption{(color online) The Fermi surface of BaFe$_{2}$As$_{2}$ for 20\%
h-doping (corresponding to Ba$_{1.6}$K$_{0.4}$Fe$_{2}$As$_{2}$, calculated 
as 40\% Cs doping in the virtual crystal approximation) \protect\cite%
{Ref1}.}
\label{FS122}
\end{figure}

If, however, we neglect this potential caveat, and concentrate on the two
best studied systems, 1111 and 122, the following relevant characteristics
can be pointed out: First, the density of states (DOS) for holes and
electrons is comparable for undoped materials; with doping, respectively one
or the other becomes dominant. For instance, for Ba$_{0.6}$K$_{0.4}$Fe$_{2}$%
As$_{2}$ the calculated DOS (in the experimental structure) for the three
hole bands varies between $1.1$ st/eV/f.u. and $1.3$ st/eV/f.u., the inner
cylinder having, naturally, the smallest DOS and the outer the largest. For
the electron bands the total DOS is $1.2$ st/eV/f.u., that is, two to three
times smaller than the total for the hole bands\cite{Ref1}. We shall see
later that this is important. Another interesting effect is that in the 122
family doping in either direction strongly reduces the dimensionality
compared to undoped compounds (in the 1111 family this effect exists, but is
much less pronounced), see Fig. \ref{FS122}. This suggests that the reason
that doping destroys the long-range magnetic order (it is believed by many
that such a destruction is prerequisite for superconductivity in
ferropnictides) is not primarily due to the change in the 2D electronic
structure, as it was initially anticipated\cite{mazin08}, but rather due to
the destruction of magnetic coupling between the layers. Indeed the most
striking difference between the undoped 1111 and undoped 122 electronic
structure is quasi two-dimensionality of the former and a more 3D character of
the latter (the difference is clear already in the paramagnetic
calculations, but 
is particularly drastic in the antiferromagnetic state), while at the
same time the observed magnetism in the 122 family is at least three times
stronger than in LaFeAsO (in the mean-field DFT calculation the difference
is quite small).

\begin{figure}[tbp]
\includegraphics[width=0.99\linewidth,viewport=40 40 560 450,clip
]{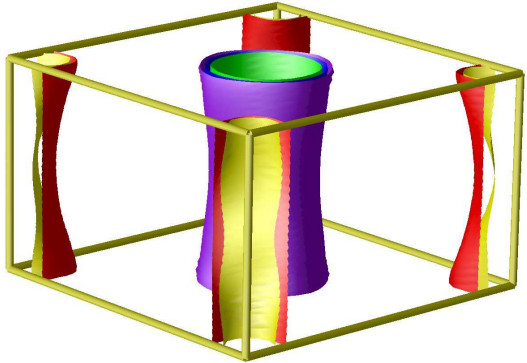}
\caption{(color online) The Fermi surface of the non-magnetic BaFe$_{2}$As$%
_{2}$ for 10\% h-doping (20\% Cs doping, virtual crysatl approximation.\protect\cite{Ref1}}
\label{FS3}
\end{figure}

The fact that the nesting is very imperfect is crucial from the point of
view of an SDW instability, making the material stable against
infinitesimally small magnetic perturbation. For superconductivity, however,
it is less important, as discussed later in the paper.


\subsubsection{Experimental evidence}

\begin{figure}[ptb]
\includegraphics[width=0.90\linewidth]{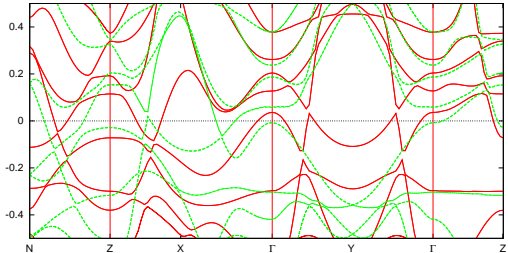}
\caption{(color online) Band structure of the orthorhombic antiferromagnetic
BaFe$_{2}$As$_{2}$ calculated for two different interlayer ordering pattern:
the experimental antiferromagnetic one (space group \#66, broken green) and the
hypothetical ferromagnetic (still antiferromagnetic in plane, space group
\#67, solid red). In both cases the magnetic moment on Fe was artifically
suppressed to 1 $\protect\mu_{B}$ by aplying a fictitious negative Hubbard
U \protect\cite{Ref1}. The point N is above the point Y.}
\label{magbands}
\end{figure}

Experimental evidence regarding the band structure and fermiology of these
materials comes, basically, from two sources: Angular resolved photoemission
spectroscopy (ARPES) and quantum oscillations measurements. The former has
an additional advantage of being capable of probing the electronic structure
in the superconducting state, assessing the amplitude and angular variation
of the superconducting gap. A potential disadvantage is that it is a surface
probe, and pnictides, especially the 122 family, are much more
three-dimensional than cuprates. This means that, first, the in-plane bands
as measured by ARPES, strongly depend on the normal momentum, $k_{\perp },$
and, second, there is a bigger danger of surface effects in the electronic
structure than in the cuprates. There are indications that the at least in
1111 compounds the surface is charged, that is to say, the doping level in
the bulk is different from that on the surface. Additionally, LDA
calculations suggest that in the magnetic prototypes, the band structure
depends substantially on interlayer magnetic ordering, again, not
surprisingly, mostly in the 122 compounds, as Fig.\ref{magbands}
illustrates. Of course, there is no guarantee that the last two layers order
in the same way as the bulk (or even with the same moment).

These caveats notwithstanding, ARPES has already provided invaluable
information. ARPES measurements have been performed for both 1111\cite%
{Liu08a,Kondo08} and 122 materials\cite{Ding08,Zhao08,Wray08,NakayamaARPES}.
These measurements demonstrated the existence of a well-defined Fermi
surface that consists of hole and electron pockets, in qualitative agreement
with the predictions of electronic structure calculations. Thus, one can say
that the topology of the Fermi surface, including the location and the
relative size of the individual Fermi surface sheets agrees with the LDA
expectation --- which is most important for the pairing models. Similarly,
it is rather clear that the ARPES bandwidth is reduced from the LDA one by a
factor of 2--2.5, similar to materials with strong itinerant magnetic
fluctuations (\textit{cf.}, for instance, Sr$_{2}$RuO$_{4}$ near a magnetic
quantum critical point\cite{Sr2RuO4}). These findings are also consistent
with the deduced normal state linear specific heat coefficient in 1111
materials (\textit{e.g., }$4-6$ mJ/mol K$^{2}$ in Ref. \cite{SH1})
corresponding to a factor 1--2 compared to the bare LDA value\cite{Note1}.
However, in the 122 compound a specific heat coefficient 63 mJ/mol K$^{2}$
was reported\cite{SH1}, to be compared with roughly 11.5mJ/mol K$^{2}$ from
the LDA calculations\cite{Ref1}. While a renormalization of 5.5 is not
consistent with either ARPES or quantum oscillations, consistency among
different experimental publications for the 122 systems is lacking as well 
\cite{SH2,SH1}.

Another experimental probe of the electronic structure is based on quantum
oscillations that measure extremal cross-section areas of the FS (ideally,
for different directions of the applied field) and the effective masses.
Such measurements are very sensitive to the sample quality, therefore so far
only a handful of results are available. However, data on the P-based 1111
compound agree reasonably well with band structure calculations\cite%
{Coldea08}, and indicate the same mass renormalization as ARPES\cite{ARPES-P}

Importantly, quantum oscillations measurements on AFM 122 compounds\cite%
{Sebastian08,Ian} indicate that even the undoped pnictides are well defined
Fermi liquids, even though a significant portion of the Fermi surface
disappears due to the opening of a magnetic gap. The frequencies of the
magneto-oscillations then suggest that the ordered magnetic state has small
Fermi surface pockets consistent with the formation of a spin-density wave.
Thus, the electronic structure of the pnictides is consistent with a
metallic state with well defined Fermi surfaces.

Besides determining the overall shape of the Fermi surface sheets, ARPES is
able to yield crucial information about the momentum dependence of the
superconducting gap. Several groups performed high quality ARPES
measurements of this effect\cite{Kondo08,Ding08,Zhao08,Wray08}. In some
cases significant differences in the size of the gap amplitude for different
Fermi surface sheets have been observed. However, there seems to be a
consensus between all ARPES groups that the gap amplitude on an individual
Fermi surface sheet depends weakly on the direction. While this seems to
favor a pairing state without nodes, one has to keep in mind that all
measurements so far have been done for fixed values of the momentum $%
k_{\perp }$, perpendicular to the planes. While it might be premature to
place too much emphesis on the relative magnitude of the gaps observed in
different bands in ARPES experiments, it is worth noting that most
experimentalists agree that in the hole-doped 122 material the inner hole
barrel and the electron barrel have comparable (and large) superconducting
gaps, while the outer hole barrel has about twice smaller gap. On the other
hand, there are first data\cite{Ding08b} indicating that in the electron
doped BaFe$_{1.85}$Co$_{0.15}$As$_{2}$ the hole and the electron bands have
about the same gap despite the hole pockets shrinking, and electron pocket
extending. Even more interesting, the most natural interpretation of the
measured fermiology is that the hole FS in BaFe$_{1.85}$Co$_{0.15}$As$_{2}$
actually corresponds to the outer ($xz/yz)$ barrel in Ba$_{0.6}$K$_{0.4}$Fe$%
_{2}$As$_{2}$ that has a small gap in that compound.

\subsubsection{Role of spin fluctuations in electronic structure}

As is clear from the above discussion, strong spin fluctuations have a
substantial effect upon the band structure. First of all, they dress
one-electron excitations providing mass renormalization, offering an
explanation for the factor 2--2.5. This is in fact a relatively modest
renormalization: it is believed that, for instance, in He$^{3}$ or in Sr$_{2}
$RuO$_{4}$ itinerant spin fluctuations provide renormalization of a factor
of 4 or larger. However, it is likely that the effect goes beyond simple
mass renormalziation. As will be discussed in detail below, there is
overwhelming evidence of large local moments on Fe, mostly from the fact
that the Fe-As bond length corresponds to a fully magnetic (large) Fe ion.
There is also evidence that the in-plane moments are rather well correlated
in the planes, and the apparent loss of the long-range ordering above $T_{N}$
is mainly due to a loss of 3D coherency between the planes\cite{3D}. It is
only natural to expect a similar situation to be true when magnetism is
suppressed by doping.

If that is the case, the electronic structure in the paramagnetic parts of
the phase diagram, at least in the vicinity of the transition, should not be
viewed as dressed nonmagnetic band, but rather as an average between the
bands corresponding to various magnetic 3D stackings (cf. Fig. \ref{magbands}%
). Fig. \ref{magbands}, corresponding to the $T=0$ magnetic moment of 1 $\mu
_{B},$ is probably exaggerating this effect, but it is still likely that in
a considerable range of temperatures and doping near the observed magnetic
phase boundary a nonmagnetic band structure is not a good starting point,
and a theory based on magnetic precursors is needed. More experiments,
particularly using diffuse scattering, and more theoretical work are needed
to clarify the issue. A discussion to this effect may be found in Ref. \cite%
{domains}. See also Section 2.3 below.

\subsection{Magnetic excitations}

\subsubsection{Experimental evidence}

Compared to cuprates and other similar compounds, two peculiarities strike
the eye. First, the parent compounds of the pnictide superconductors assume
an antiferromagnetic structure, where neighboring Fe moments are parallel
along one direction  withinin the FeAs plane and antiparallel along the
other. Neutron scattering data yield ordered moments per Fe of $0.35\mu _{B}$
for LaFeAsO\cite{neutronsLa}, $0.25\mu _{B}$ for NdFeAsO\cite{neutronsNd}, $%
0.8\mu _{B}$ for CeFeAsO\cite{neutronsCe}, and $0.9$ $\mu _{B}$ for BaFe$_{2}
$As$_{2}$\cite{neutronsBa}. Intriguingly, in NdFeAsO the ordered moment at
very low temperatures increases by a factor of 3 to 4 at the temperature
corresponding to the ordering of Nd-spins\cite{neutronslowTNd}. Note that
the correct magnetic structure has been theoretically predicted by DFT
calculations\cite{mazin08,Fang}, which, moreover, consistently overestimated
the tendency to magnetism (as opposed to the cuprates). Second, the
magnetically ordered state remains metallic. As opposed to cuprates or other
transition metal oxides, the undoped systems exhibit a small but well
established Drude conductivity\cite{Hu08}, display magneto-oscillations\cite%
{Sebastian08} and have Fermi surface sheets of a partially gapped metallic
antiferromagnetic state\cite{Hsieh08}. Above the magnetic ordering
temperature a sizable Drude weight, not untypical for an almost semimetal
has been observed. Further, the ordered Fe magnetic moment in the 1111
systems depends sensitively on the rare earth ion, very different from YBa$%
_{2}$Cu$_{3}$O$_{6}$ where yttrium can be substituted by various rare earth
elements with hardly any effect on the Cu moment. Note that the rare earth
sites project onto the centers of the Fe plaquettes and thus do not
exchange-couple with the latter by symmetry. Finally, the magnetic
susceptibility of BaFe$_{2}$As$_{2}$ single crystals\cite{Wang08} above the
magnetic transition shows no sign for an uncoupled local moment behavior.

\subsubsection{Itinerant versus local magnetism}

The vicinity of superconductivity to a magnetically ordered state is the key
motivation to consider pairing mechanisms in the doped systems that are
linked to magnetic degrees of freedom. Similar to cuprate superconductors,
proposals for magnetic pairing range from quantum spin fluctuations of
localized magnetic moments to fluctuations of paramagnons as expected in
itinerant electron systems. To judge whether the magnetism of the parent
compounds is localized or itinerant (or located in the crossover regime
between these two extremes) is therefore crucial for the development of the
correct description of magnetic excitations and possibly the pairing
interactions in the doped systems.

In our view the case at hand is different from such extreme cases as undoped
cuprate on one end and weak itinerant magnets like ZrZn$_{2}$ on the other.
While being metals with partially gapped Fermi surface, there is evidence
that Fe ions are in a strongly magnetic states with strong Hund rule
coupling for Fe. This results in a large magnetic moment --- but only for
some particular ordering patterns (for comparison, in FeO and similar
materials LDA produce large magnetic moment regardless of the imposed long
range order). While it is obvious that ferropnictides are not Mott
insulators with localized spins, interacting solely with near neighbors, a
noninteracting electron system may be not a perfect starting approximation
either. To make progress we have to decide what is the lesser of two evils
and use it, even realizing the problems with the selected approach. Given
the above mentioned experimental facts, our preference is that these systems
are still on the itinerant side.

A feature that has attracted much interest is the quasi-nesting between the
electron and the hole pockets. The word \textquotedblleft
quasi\textquotedblright\ is instrumental here: even the arguably most nested
undoped LaFeAsO is very far from the ideal nesting and even worse in the
(more magnetic) BaFe$_{2}$As$_{2}.$ Indeed, it has been observed that in the
LDA calculations the nonmagnetic structure in either compound is stable with
respect to an infinitesimally small AFM perturbations, but strongly unstable
with respect to finite amplitude perturbations. This can be understood from
the point of view of the Stoner theory, applied to a finite wave vector 
\textbf{Q}: the renormalized static spin susceptibility (in the DFT the RPA
approximation is formally exact) can be written as%
\begin{equation}
\chi _{LDA}(\mathbf{Q)}=\frac{\chi _{0}(\mathbf{Q})}{1-I\chi _{0}(\mathbf{Q})%
},  \label{RPA}
\end{equation}%
where $I$ is the Stoner factor of iron, measuring the intra-atomic Hund
interaction (in the DFT, it is defined by the second variation of the
exchange-correlation functional with respect to the spin density). While the
denominator in Eq. \ref{RPA} provides a strong enhancement of $\chi $,
albeit not exactly at $\mathbf{Q=}(\pi ,\pi )$, but at a range of the wave
vectors near $\mathbf{Q}$), it does not by itself generate an instability.
One can say that an infinitesimally weak magnetization can only open a gap
over a very small fraction of the Fermi surface. However, a large-amplitude
spin density wave opens a gap of the order of the exchange splitting, $IM$,
where $M$ is the magnetic moment on iron, and, obviously, affects most of
the conducting electrons. In other words, the magnetism itself is generated
by the strong Hund rule coupling on Fe (just as in the metal iron), but the
topology of the Fermi surface helps select the right ordering pattern.
Formation of the magnetic moments is local; arranging them into a particular
pattern is itinerant.

There are several corollaries of this fact that are important for pairing
and superconductivity. First, despite the fact that the overall physics of
these materials is more on the itinerant side than on the localized side
(see a discussion to this effect later in the paper), it is more appropriate
to consider magnetic moments on Fe as local rather than itinerant (as for
instance in the classical spin-Peierls theory). Note that the same is true
for the metal iron as well. Second, the \textit{interaction} among these
moments is \textit{not} local, as for instance in superexchange systems (it
appears impossible to map the energetics of the DFT calculations onto a two
nearest neighbor Heisenberg model\cite{Yares}). The AFM vector is not
determined by local interactions in real space (as for instance in the $%
J_{1}+J_{2}$ models, see below), but by the underlying electronic structure
in reciprocal space. Third, since the energy gain due to formation of the
SDW mainly occurs at finite (and large, $IM$ is on the order of eV)
energies, looking solely at the FS may be misleading. Indeed, FeTe is one
compound where the Fe moments apparently do not order into a $\mathbf{Q=}%
(\pi ,\pi )$ SDW, but in a more complex structure corresponding to a
different ordering vector\cite{FeTe}, despite the fact that the FS shows
about the same degree of nesting (Fig.\ref{FeTe}) as LaFeAsO and a
noticeably better nesting than BaFe$_{2}$As$_{2}.$ DFT calculations
correctly identify the ground state in all these cases, and the origin can
be traced down again to the opening of a partial gap: in both 1111 and 122
compounds the $\mathbf{Q=}(\pi ,\pi )$ is about the only pattern that opens
such a gap around the Fermi level, while in FeTe comparable pseudogaps open
in both magnetic structures (and the calculated energies are very close, the
actual experimental structure being slightly lower\cite{Ref2}).

\subsubsection{Perturbative itinerant approach}

Even if one accepts the point of view that the magnetism in the Fe-pnictides
is predominantly itinerant, the development of an adequate theory for the
magnetic fluctuation spectrum is still highly nontrivial. As pointed out
above, there are strong arguments that the driving force for magnetism is
not Fermi surface nesting but rather a significant local Hund's and exchange
coupling. This can be quantitatively described in terms of a multiband
Hubbard type interaction of the Fe-$3d$ states 
\begin{align}
H_{int}& =U\sum_{i,a}n_{ia\uparrow }n_{ia\downarrow }+U^{\prime
}\sum_{i,a>b}n_{ia}n_{ib}  \notag \\
& -J_{H}\sum_{i,a>b}\left( 2\mathbf{s}_{ia}\cdot \mathbf{s}_{ib}+\frac{1}{2}%
n_{ia}n_{ib}\right)   \notag \\
& +J\sum_{i,a>b,\sigma }d_{ia\sigma }^{\dagger }d_{ia\overline{\sigma }%
}^{\dagger }d_{ib\overline{\sigma }}d_{ib\sigma },
\end{align}%
with intra- and inter-orbital Coulomb interaction $U$ and $U^{\prime }$,
Hund's coupling $J_{H}$ and exchange coupling $J$, respectively. Here $a$, $b
$ refer to the orbitals in a Wannier type orbital at site $i$. $X$-ray
absorption spectroscopy measurements support large values for the Hund's
couplings that lead to a preferred high spin configuration,\cite{Kroll08}
leading to larger values of $J_{H}$. The importance of the Hund coupling for
the normal state behavior of the pnictides was recently stressed in Ref.\cite%
{Kotliar}.

Weak coupling expansions in these interaction parameters may not capture
quantitative aspects of the magnetism in the pnictides. Nevertheless, it is
instructive to summarize the main finding of the result of weak coupling
expansions, in particular as they demonstrate the very interesting and
nontrivial aspects that results from interband interactions with almost
nested hole and electron Fermi-surfaces\cite{LeeRG,ChubukovRG,Zlatko}. For
an ideal semimetal (two identical hole and electron bands with the Fermi
energies $E_{h}$ and $E_{e})$ all susceptibilities at the nesting vector 
\textbf{Q} diverge as $\log |E_{h}/E_{e}-1|$. Depending on the details of
electron-electron interaction this signals an instability, at $E_{h}=E_{e},$
to a spin density wave state or to a superconducting state for infinitesimal
interaction. The corresponding interference between particle-hole and
particle-particle scattering events can be analyzed by using a
renormalization group approach. For $J_{H}=J=0$, the authors of Ref.\cite%
{ChubukovRG} find that at low energies the interactions are dominated by
Cooper pair-hopping between the two bands, favoring an $s^{\pm }$%
-superconducting state that is fully gapped on each Fermi surface sheet, but
with opposite sign on the two sheets. It is worth pointing out that this
pairing mechanism is due to very generic interband scattering, not
necessarily due to \emph{spin-fluctuations}, as all particle-hole and
particle-particle scattering events enter in essentially the same matter. An 
$s^{\pm }$-state was also obtained using a functional renormalization group
approach\cite{LeeRG}, where the authors argue that the pairing mechanism is
due to collective spin fluctuations that generate a pairing interaction at
low energies. The appeal of these calculations is clearly that controlled
and thus robust conclusions can be drawn. On the other hand, as discussed
below, the Fermi surface nesting is less crucial as is implied by these
calculations.

Attempts to include sizable electron-electron interactions within an
itinerant electron theory are based on the partial summation of ladder and
bubble diagrams, in the spirit of Eq.\ref{RPA}. This leads to the RPA type
theory of Ref.\cite{Kuroki08,Graser08,Graser09,Scalapino} and the
fluctuation exchange approximation of multiband systems\cite{Yao,Sknepnek}.
RPA calculations yield a magnetic susceptibility that is peaked at or near $%
\mathbf{Q=}\left( \pi ,\pi \right) $. For parameters where the Fermi surface
around $\Gamma $ is present, the dominant pairing channel is again the $%
s^{\pm }$-state, while $d$-wave pairing occurs as one artificially
eliminates this sheet of the Fermi surface. The exchange of paramagnons
between Fermi surface sheets is shown to be an efficient mechanism for spin
fluctuation induced pairing. The fluctuation exchange (FLEX) approach is to
some extent a self consistent version of the RPA theory\cite{Bickers89}.
While the method is not very reliable to address high energy features, the
description of the low energy dynamics spin response, the low energy
electronic band renormalization and, the nature of the pairing
instabilityare  rather reliable. The fact that several orbitals matter in
the FeAs systems is also of help as FLEX type approaches can be formulated
as theories that become exact in the limit of large fermion flavor\cite%
{Abanov}. Refs.\cite{Yao,Sknepnek} performed FLEX calculations for the FeAs
systems and find once again that the dominant pairing state is an $s^{\pm }$%
-state, even though Ref.\cite{Yao} also find a $d$-wave state in a regime
where the magnetic fluctuation spectrum is peaked at vectors away from $%
\mathbf{Q=}\left( \pi ,\pi \right) $. These authors find a solution that is
numerically close to a compact form 
\begin{equation}
\Delta \left( \mathbf{k}\right) =\Delta _{0}\cos (ak_{x})\cos (ak_{y}),
\label{s+-}
\end{equation}%
but this form is neither required by symmetry nor can be consistently
deduced from any low-energy theory (where pairing occurs at or near the
Fermi surface). We will come back to this issue later in this review.

To summarize, numerous calculations that start from an itinerant description
of the magnetic interactions yield an $s^{\pm}$ pairing state caused by the
exchange of collective interband scattering or paramagnons.

\subsubsection{J$_{1}$-J$_{2}$ model}

The initially assumed (although later refuted by the experiment\cite{Wang})
absence of the Drude weight in undoped ferropnictides
has been taken as evidence for the fact that they are in
the vicinity of a Mott transition and should be considered as bad metals
with significant incoherent excitations\cite{Si08}. If correct, it is clearly
appropriate to start from a theory of localized spins, analogous to what is
believed to be correct in the cuprate superconductors\cite%
{Kivelson08,Subir08} (it is worth noting that proximity to a Mott 
transition is a sufficient, but not necessary condition for existence
of local moments). If the dominant magnetic interactions are between
nearest and next nearest neighbor Fe-spins, the following  model describes 
the localized spins:
\begin{equation}
H=J_{1}\sum_{\left\langle i,j\right\rangle }\mathbf{S}_{i}\cdot \mathbf{S}%
_{j}+J_{2}\sum_{\left\langle \left\langle i,j\right\rangle \right\rangle }%
\mathbf{S}_{i}\cdot \mathbf{S}_{j}  \label{J1J2}
\end{equation}%
Here, $J_{1}$ and $J_{2}$ are the superexchange interactions between two
nearest-neighbor and next-nearest-neighbor Fe sites, respectively. A
geometrical argument can be made\cite{Taner,Si08} that indeed the two
superexchange paths $via$ As have comparable strength (however, this
argument fails to recognize that the direct overlap between Fe orbitals in
pnictides is very large\cite{SinghDu}, thus leading to a strong enhancement
of the nearest neighbor antiferromagnetic exchange in the localized picture%
\cite{Li}, and that in metals superexchange is not the only and usually not
the most important magnetic interaction). When $J_{1}>2J_{2}$ the
conventional Neel state has the lowest energy, when $J_{1}<2J_{2}$ the stripe
order emerging in the experiment is the lowest magnetic state. The system is
frustrated if $J_{1}=2J_{2}.$

Upon doping the poor metal (strictly the insulator) described by Eq. \ref%
{J1J2} with charge carriers can be investigated for superconductivity, with
pairing stabilized by strong quantum spin fluctuations. In Ref.\cite%
{Sachdev02} a single band of carriers was investigated leading to either $%
d_{x^{2}-y^{2}}+id_{xy}$ or $d_{xy}$-pairing, depending on the carrier
concentration and the precise ratio of $J_{1}$ and $J_{2}$. A more realistic
theory for the pairing in the $J_{1}$-$J_{2}$ model in the pnictides must of
course include at least two bands and was developed in Ref.\cite{Bernevig}.
For sufficiently large $J_{2}$, the $s^{\pm }$-state is once again the
dominating pairing state. It may seem strange that this strong coupling
theory based upon the (unlikely, from the experimental point of view)
proximity to a Mott transition has essentially the same pairing solutions ($d
$-wave for one Fermi surface sheet and $s^{\pm }$-wave for two Fermi surface
sheets separated by $\mathbf{Q}$), as the RPA calculation of \cite{Kuroki08}%
. In Section 3 we will explain that this is not surprising at all and that
even a totally unphysical theory may lead to perfectly sensible results for
superconductivity, as long as it has the same structure of magnetic
excitations in the reciprocal space.

\subsection{Magneto-elastic coupling}

The parent compounds exhibit a structural and a magnetic transition,
strongly suggesting that magnetoelastic coupling plays a role in the physics
of pnictides in general and in superconductivity in particular. Electronic
structure calculations for a non-magnetic state indicate that the
electron-phonon interaction in the pnictides is rather modest and definitely
not sufficient to explain superconducting transition temperatures of $50$ K%
\cite{Boeri08,mazin08}. However, as these calculations were based on the
nonmagnetic electronic structure, effects of local magnetism on iron were
entirely neglected. Indeed, the equilibrium position of As calculated under
this assumption are quite incorrect and the force constant for the Fe-As
bond is $30\%$ higher than it should be. On the other hand, fully magnetic
AFM calculations, while overestimating the ordered moment, produce highly
accurate equilibrium structures and the force constant in agreement with
experiment\cite{domains}. It was pointed out that including soft magnetism
in the calculation, i.e. magnetism with directional and amplitude
fluctuations, may substantially enhance the electron-phonon coupling\cite%
{Yndurain}. The emphasis is on \textquotedblleft soft\textquotedblright\ :
additional reduction of the force constants of the Fe-As bonds does not come
from the fact that the moment exists, but from the fact that the amplitude
of the moment depends on the bond length. Intriguingly, in the 1111 systems
the AFM transition occurs somewhat below a structural phase transition. Both
transitions seem to be of the second order, or of very weakly first order%
\cite{Mandrus}. In 122 compounds the structural and magnetic orders emerge
simultaneously through a strong first order transition\cite{Rotter08b,NiNi}.

In the ordered state, Fe spins are parallel along one direction and
antiparallel along the other. Since we expect the bond length for parallel
and antiparallel Fe-spin polarization to be distinct, magnetism couples
strongly to the shear strain $\varepsilon _{\mathrm{shear}}=\varepsilon
_{xy}-\varepsilon _{yx}$. Thus, $\varepsilon _{\mathrm{shear}}\neq 0$ should
invariably occur below the Neel temperature. Experiment finds that the
ferromagnetic bonds are shorter than antiferromagnetic bonds. From the point
of view of superexchange interaction it seems somewhat surprising that
ferromagnetic bonds shorten and the superexchange-satisfied bonds expand.
Yet this behavior is exactly the same as the DFT calculations had predicted\cite{Taner},
and it can
be traced down to one-electron energy (the observed sign of the orthorhombic
distortion simply lowers the one-electron DOS at the Fermi level)\cite{tbp}.

What remains puzzling is however why in the 1111 family the structural
transition occurs above $T_{N}.$ Naively, this fact could be taken as
evidence for a hypothesis that elastic degrees of freedom are the driving
force and that magnetism is secondary. There are strong quantitative and
qualitative arguments against this view. First, numerous DFT calculations%
\cite{Rosner,Samolyk,domains} converge to the correct orthorhombic structure
(with correct sign and magnitude of the distortion), if performed with AFM
magnetic ordering, and to a tetragonal solution if done without magnetism.
On the other hand, the antiferromagnetism is obtained even without allowing
for a structural distortion. In other words, magnetism is essential for the
distortion, but the distortion is not needed for the magnetism.

\begin{figure}[tbp]
\includegraphics[width=0.95\linewidth]{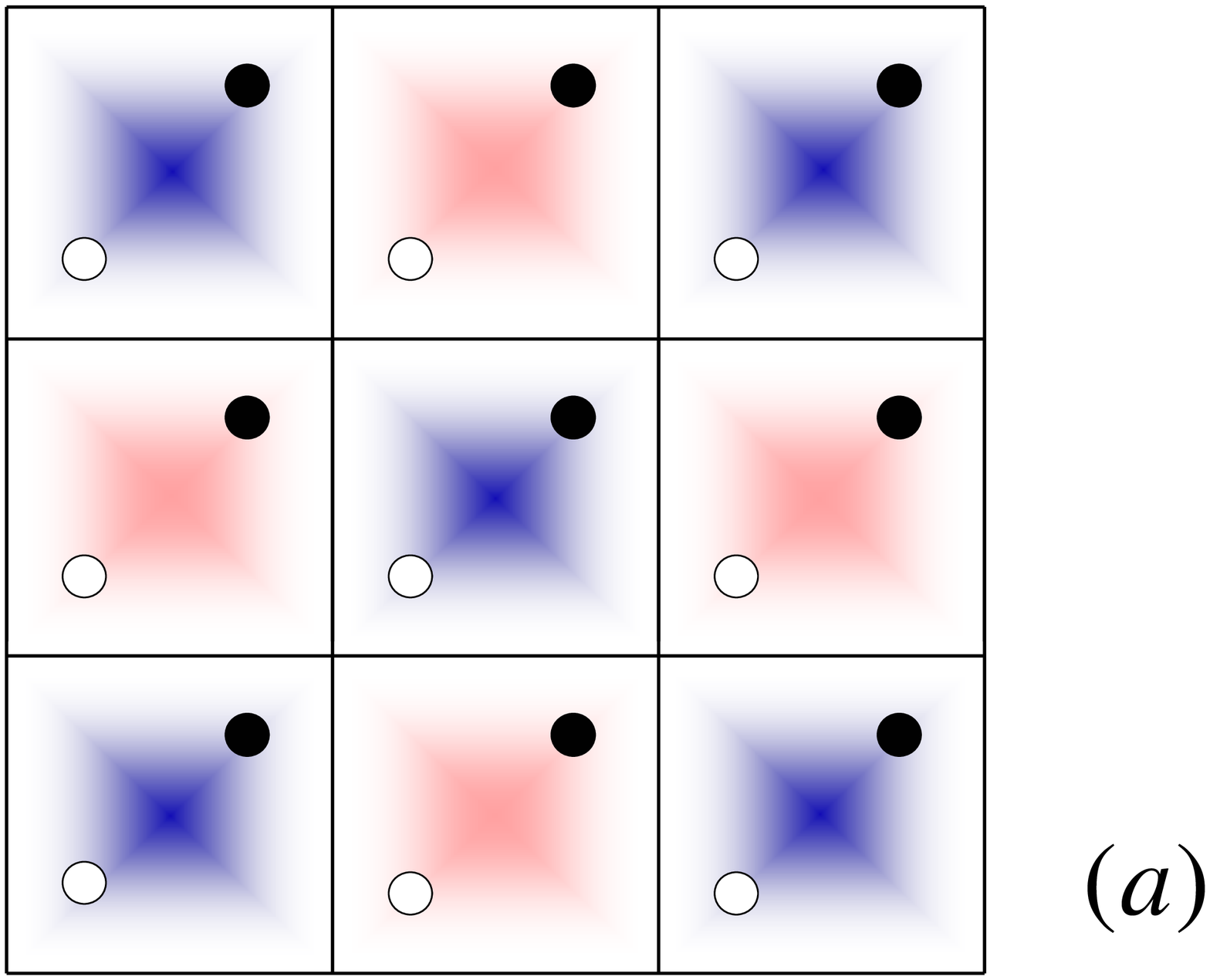}
\includegraphics[width=0.95\linewidth]{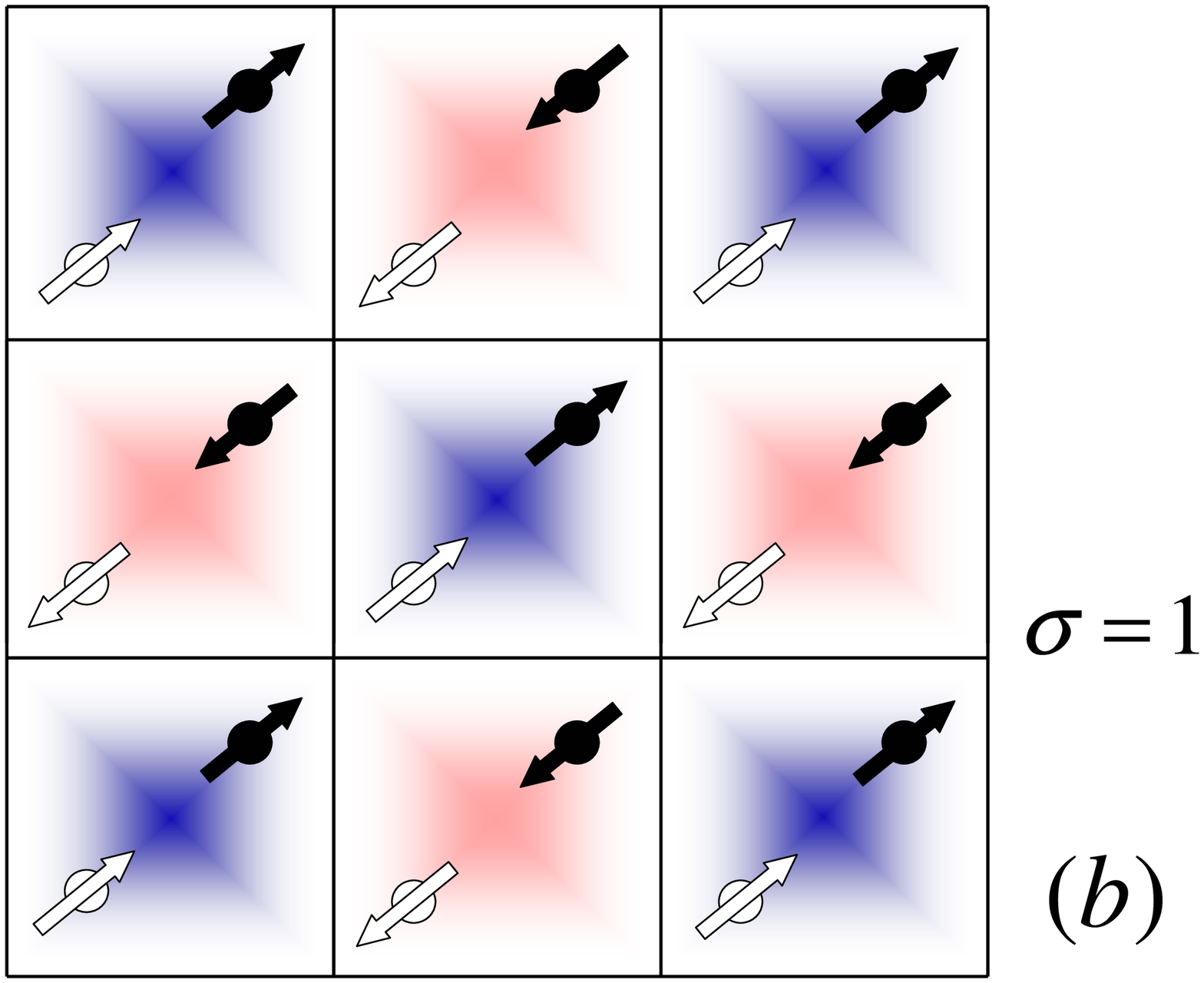}
\includegraphics[width=0.95\linewidth]{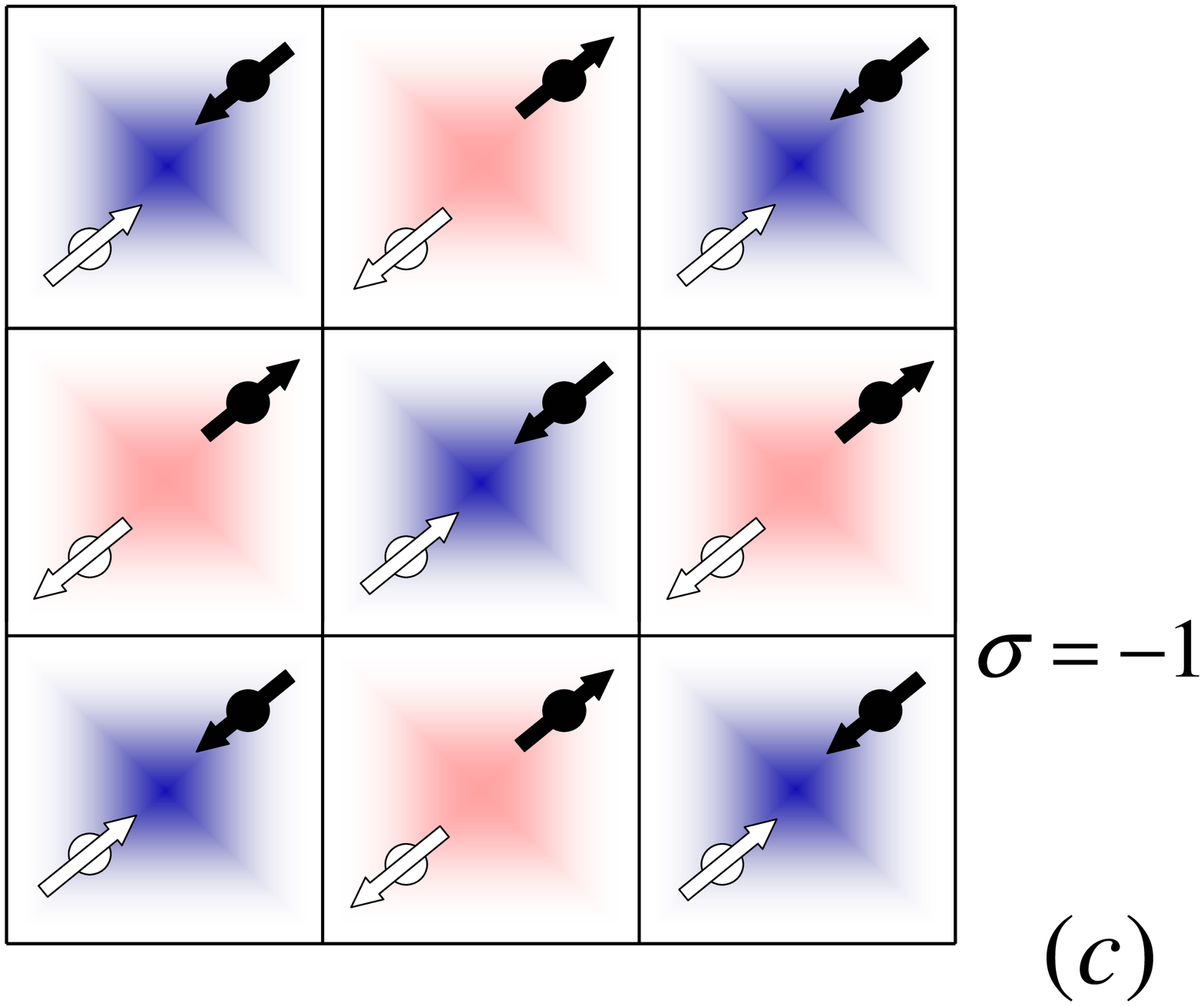}
\caption{(color online) (a)Fe$_{2}$ lattice with the fully symmetric unit
cells shown. The full circles denote one sublattice, the hollow ones the
other. Shading shows ordering corersponding to the vector $\mathbf{Q=}\left( \protect\pi %
,\protect\pi \right)$ in the Fe$_{2}$ lattice; for each ssublattice,
spins in the pink unit cells are opposite to the spins in the blue cells, but relative
orientation of the two sublattices is arbitrary.
  (b) Ordered state with $\mathbf{Q=}\left( \protect\pi ,\protect\pi %
\right) $ and with parallel orientation of the spins in the unit cell ($%
\protect\sigma =1$). (c) Same ordering vector $\mathbf{Q=}\left( \protect\pi %
,\protect\pi \right) $, but with antiparallel orientation of the spins in
the unit cell ($\protect\sigma=-1$). }
\label{sig}
\end{figure}

There exists also a very general argument that demonstrates that the
magnetism is indeed primary and the structural distortion secondary.
Historically the relevant physics was first encountered in the 2D $J_{1}$-$%
J_{2}$ model\cite{Premi90}, and applied to ferropnictides in Refs.\cite%
{Kivelson08,Subir08}. Below we will reformulate this argument form a general
point of view. We begin with a unit cell that contains two Fe sites (just as
the actual cristallographic unit cell for the FeAs trilayer). The most
natural choice of the origin is in the middle between these two Fe cites
(Fig. \ref{sig}a ). The coordinates of the atoms are $\mathbf{r}_{ij}^{+}=%
\mathbf{R}_{ij}+\mathbf{d}$, $\mathbf{r}_{ij}^{-}=\mathbf{R}_{ij}-\mathbf{d,}
$ $\mathbf{d}=(\frac{1}{4},\frac{1}{4}),$ where $\mathbf{R}_{ij}$ ($i,j$
integer) are the coordinates of the centers of the unit cells. This
naturally implies partitioning the entire lattice into two sublattives,
shown as open and solid dots in Fig. \ref{sig}a.

Both ferro- and antiferromagnetic checkerboard orderings correspond to a $%
\mathbf{Q}=(0,0)$ perturbation of the uniform state, since in both cases all
unit cells remain identical. The Fourier transform of either patter contains
only momenta corresponding to the reciprocal lattice vectors.  Conversely, a
spin density wave with the quasi-momentum $\mathbf{Q}=(\pi ,\pi )$
corresponds to flipping all spins in every other unit cell, as illustrated
in Fig. \ref{sig}b,c by shading colors (blue cells have the magnetization
density opposite to that of the pink cells). It is evident from Fig. \ref%
{sig}b and c that this imposes no requirement upon the mutual orientation of
the two sublattices. Again, one can say that the susceptibility as a
function of \textit{quasi}momentum $\mathbf{q}$ inside the first Brillouin
zone does not describe fluctuations of the magnetic moment of two ions in
the same unit cell with respect to each other, for that purpose one needs to
know the linear response at all momenta $\mathbf{q+G}$, where $\mathbf{G}$
is an arbitrary reciprocal lattice vector.

Let us assume that the most stable mean field phase corresponds to N\'{e}el
order in each of the two sublattices. In the $J_{1}$-$J_{2}$ language that
corresponds to $J_{2}>J_{1}/2,$ in the itinerant language to an instability
in $\chi $ at $\mathbf{Q}=(\pi ,\pi )$. Moreover, it is obvious from Fig. %
\ref{sig}b,c that in the classical ground state one sublattice does not
exchange-couple at all to the other, so the classical ground state is
infinitely degenerate. this is however not important for the following
discussion, what matters is that the two extreme cases are always
degenerate, the one where two spin in the same cell are parallel (Fig. \ref%
{sig}b) or antiparallel (Fig. \ref{sig}c). In the $J_1+J_2$ model the infinite
degeneracy is reduced by quantum fluctuations, but the double degeneracy
remains, while in the LDA it is only double degenerate already on the
mean-field level\cite{Yarnote}.

It is instructive \cite{Premi90} to introduce two order parameters
corresponding to the Neel (checkerboard) ordering for each sublattice, $%
\mathbf{m}_{\pm }=\sum_{ij}(-1)^{i+j}\mathbf{M}_{ij}^{\pm },$where $\mathbf{M%
}_{ij}^{\pm }$ are the magnetic moments of the two Fe's in the unit cell $ij.
$ Following Ref. \cite{Premi90} one can introduce the third (scalar) order
parameter, $\sigma =\sum_{ij}\sigma _{ij}=\sum_{ij}\mathbf{M}_{ij}^{+}\cdot 
\mathbf{M}_{ij}^{-}$. Now $\sigma >0$ corresponds to parallel orientation of
the magnetization inside the unit cell (Fig. \ref{sig}b) while $\sigma <0$
refers to antiparallel orientation (Fig. \ref{sig}c). In the former case $%
\sigma >0$, neighboring Fe spins are parallel along the diagonal and
antiparallel along the counter-diagonal. The situation is reversed for $%
\sigma <0$. These two configurations are degenerate and correspond to the
frequently discussed 'stripe' magnetic order. In two dimensions, according
to the Mermin-Wagner theorem, $\sigma $ is the only order parameter that can
be finite at finite temperature. Therefore the presumably largest energy
scale of the system, the mean field transition temperature of each
sublattice, $T^{\ast }$ ($\sim J_{2}$ in the local model, and the energy
difference $E_{FM}-E_{AFM}$ in the itinerant picture), does not generate any
phase transition, but rather starts a crossover regime where the correlation
length $\xi _{m}$ for the $\mathbf{m}_{\pm }$ order parameter becomes much
longer that the lattice parameter.

In this regime, one can investigate a possibility of a phase transition
corresponding to the $\sigma $ order parameter. It is important to realize
that $\sigma $ does not have to change sign along a domain wall of the
magnetization. This ensures that $\sigma $ can order even though the
sublattice magnetization vanishes. $\sigma $ does couple to the (long-range)
fluctuations of $\mathbf{m;}$ integrating these fluctuations out one will
obtain an effective Hamiltonian coupling $\sigma _{ij}$ and $\sigma
_{i^{\prime }j^{\prime }}$ as far as $\xi _{m},$ meaning that even very
small coupling between $\mathbf{m}_{+}$ and $\mathbf{m}_{-}$ will produce a
phase transition to a finite $\sigma $ at a temperature $T_{s}\sim J_{1}\xi
_{m}^{2}(T_{s})\sim J_{1}\exp (J_{2}/T_{s})$. Solving this for $T_{s}$, one
gets $T_{S}\sim J_{2}/\log (J_{2}/J_{1})$. Note that here again $J_{1}$ and $%
J_{2}\sim T^{\ast }$ just characterize the relevant energy scales and by no
means require the validity of the $J_{1}+J_{2}$ model.

\begin{figure}[tbp]
\includegraphics[width=0.99\linewidth
]{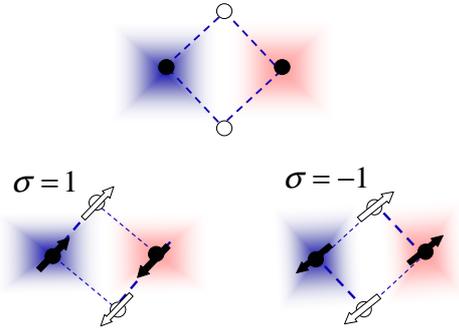}
\caption{(color online) \ Magnetoelastic coupling: The two atoms per unit
cell are denoted by filled and open circles. A ferromagnetic bond leads to a
shortening of the nearest neighbor lattice constant (bold dashed lines), while an
antiferromagnetic bond leads to a longer lattice constanti (thin dashed lines). Depending on the
relative orientation of the two sublattices (i.e. the sign of $\protect%
\sigma $), two distortions \ with opposite sign of $\protect\varepsilon _{%
\mathrm{shear}}$ are possible. }
\label{struct}
\end{figure}

As mentioned above $\sigma $ is positive (negative) for ferromagnetic
(antiferromagnetic) bonds, see Fig.\ref{struct}. Thus $\sigma $ couples
bilinearly to the order parameter of the orthorhombic structural transition%
\begin{equation}
F_{c}=\gamma \varepsilon _{\mathrm{shear}}\sigma .
\end{equation}%
When the expectation value of $\sigma $ is nonzero below a transition
temperature $T_{s}$, the tetragonal symmetry is spontaneously broken leading
to $\varepsilon _{\mathrm{shear}}\neq 0$. We see that $T_{s}$ is suppressed
from $T^{\ast }$ rather weakly (logarithmically) and that even a weak
coupling between the two sublattices would produce a structural phase
transition.

The third energy scale existing in the problem is set by the interlayer
magnetic coupling, $J_{\perp }.$ In the DFT we found $J_{\perp }\lesssim 1$
meV in LaFeAsO and $J_{\perp }\sim 16$ meV in BaFe$_{2}$As$_{2}$\cite{Ref1}.
This huge difference defines the different behavior of these two compounds.
In the former the Neel transition temperature for a sublattice ordering is
on the order of $T^{\ast }/\log (T^{\ast }/J_{\perp }),$ logarithmically
smaller than $T_{s},$ while in the latter one expects a much larger $T_{N}$,
and likely larger than the $T_{s}$ for an individual FeAs plane.

The phase between $T_{N}$ and $T_{s},$ if $T_{s}>T_{N},$ was dubbed
\textquotedblleft nematic\textquotedblright\ in Refs. \cite%
{Kivelson08,Subir08}, as the order parameter $\left\langle \sigma
\right\rangle \neq 0$ even though $\left\langle \mathbf{M}_{ij}\right\rangle
=0$, as expected for an axial, as opposed to vectorial order parameter. The 
first order nature of the transition in the 122 systems is then likely a
consequence of the coupling to soft elastic degrees of freedom, and/or of
nonlinear interactions. A more rigorous treatment of the described physics
will be published elsewhere\cite{Fernandes}. There is another interesting
experimental evidence for the unconventional nature of the magneto-elastic
coupling in these systems. In the 122 systems the structural distortion $%
\propto \varepsilon _{\mathrm{shear}}$ and the sublattice magnetization seem
to be proportional to each other.\cite{Geibel} At a second order transition,
symmetry arguments imply however that the former should be proportional to
the square of the sublattice magnetization. At a first order transition, no
such strict connection can be established, however one expects that the
generic behavior is recovered as the strength of the first order transition
gets smaller, realizable via alcaline earth substitution. Experiments show
that the mentioned linear behavior is similar for Ca, Ba or Sr\cite{Geibel2}%
. In our view this behavior is evidence for the fact that the first order
transition in the 122 systems is never close to being weak. Arguments that
the first order character of the magneto-elastic phase transition originates
from the lattice instabilities near the onset of spin-density wave order
were recently given in Ref.\cite{Gorkov08}. However, further discussion
clearly goes beyond the limit of this review.

The fact that at the structural transition (and even above), magnetic
correlations in plane are already well established, with large correlation
lengths, explains many otherwise mysterious observations. A more detailed
discussion can be found in Ref. \cite{domains}.

This picture is not without ramifications for superconductivity. First and
foremost, it implies that at superconducting composition ferropnictides,
especially the 1111 family, are not really paramagnetic, bat rather systems
with a large in-plane magnetic correlation length, much larger than the
lattice parameter and likely much larger than the superconducting
correlation length. Second, the excitation structure in such a system is
unusual and cannot be entirely described in terms of $\chi (\mathbf{Q),}$
where $\mathbf{Q}=(\pi ,\pi ),$ since such a description loses the physics
associated with the parameter $\sigma .$ Finally, it implies that the
lattice and spin degrees of freedom do not fluctuate independently and are
naturally connected to each other. Therefore a detailed quantitative theory
for the pairing state will have to include lattice vibrations. Conversely,
experiments that find evidence for a lattice contribution to the pairing
mechanism should not be considered as evidence against magnetic pairing.

\subsection{Other excitations}

While everybody's attention is attracted to magnetic pairing mechanisms and
spin fluctuations, it would be premature and preposterous to exclude any
other excitations from consideration. First of all, it might be still too
early to discard the venerable phonons. While there is no question that the
calculations performed so far \cite{Boeri08,mazin08} were accurate and the
linear response technique used had proved very reliable before (MgB$_{2},$
CaC$_{6}$ $etc.),$ these calculation by definition do not take into account
any effects of the magnetism. As discussed above, it is very likely that the
ground state even in the so-called nonmagnetic region of the phase diagram
is characterized by an AFM correlation length long enough compared to the
inverse Fermi vector. In this case, the amplitude of the magnetic moment of
Fe (even though its direction fluctuates in time) is nonzero and the
electronic structure is sensitive to it. Calculations suggest that a phonon
stretching the Fe-As bond will strongly modulate this magnetic moment and
thus affect the electronic structure at the Fermi level more than for a
nonmagnetic compound (or, for that matter, a magnetic compound with a hard
magnetic moment). Softness of the Fe moments, variationally, provides an
additional route for electron-phonon coupling and should therefore always
enhance the overall coupling constant. Whether this is a weak or a strong
effect, and whether the resulting coupling is stronger in the intraband
channel (enhancing the $s_{\pm }$ superconductivity) or in the interband
channel (with the opposite effect), is an open question. Only preliminary
results are available\cite{Yndurain}.

Besides the phonons and the spin fluctuation, charge (polarization)
fluctuations can also, in principle, be pairing agents. To the great
surprise of the current authors, nobody has yet suggested an acoustic
plasmon mechanism for ferropnictides, a mechanism that was unsuccessfully
proposed for cuprates, for MgB$_{2}$ and for CaC$_{6}.$ Presumably the
apparent lack of strong transport anisotropy in 122 and the absence of
carriers with largely disparate mass prevented these usual suspects from
being discussed.

It is not only the harsh condition on the very existence of acoustic
plasmons, but a very general malady (better known in the case of acoustic
plasmons, but generally existing for any sort of exciton pairing) that
prevents plasmonic superconductivity in most realistic cases: lattice
stability. Basically, efficient pairing of electrons via charge excitations
of electronic origin requires overscreening of electrostatic repulsion ---
which by itself does not constitute a problem. But since the ion-ion
interaction is screened by the same polarization operator as
electron-electron interaction, there is an imminent danger that the former
is overscreened as well. This is an oversimplified picture
(electron-electron susceptibility differs from the response to an external
field on the level of vertex corrections), but it captures the essential
physics.

This danger was appreciated by the early proponents of the excitonic
superconductivity, W. Little\cite{Little} and V. Ginzburg\cite{Ginzburg},
therefore they proposed space separation between a highly polarizable
insulating media, providing excitons, and a metallic layer or string where
the superconducting electrons live. The sandwich structure of the As-Fe-As
trilayer reminds us of the Ginzburg's \textquotedblleft
sandwich\textquotedblright\ (\textquotedblleft Ginzburger\textquotedblright\
) and tempts to revisit his old proposal.

This was done recently by Sawatzky and collaborators\cite{Sawatzky} who
pointed out that As is a large ion (Pauling radius for As$^{4-}$ is 2.2 \AA %
) and ionic polarizability grows with the radius cube. Since the conducting
electrons are predominantly of Fe origin, they suggested pairing of Fe d
electrons $via$ polarization of As ions. So far, this proposal was received
with a skepticism that can be summarized as follows. (1) Analyzing the
muffin-tin projected character of the valence bands, as it was done in Ref. 
\cite{Sawatzky} is generally considered to be an unreliable way to estimate
the hybridization between different ions; indeed the largest part of the
electronic wave function refers to the interstitial space, which is
naturally identified as mostly As-like. (2) Removal of the As orbitals from
the basis leads to a strong reduction of the valence band width, indicating
that hybridization between Fe and As is about as strong as direct Fe-Fe
hopping. (3) When Bloch functions are projected upon the Fe-only Wannier
functions, the latter come out very diffuse and extend way beyond the Fe
ionic radius. That is to say, negligible hybridization between Fe and As,
that is prerequisite for the scenario promoted in Ref. \cite{Sawatzky},
appears to be a rather questionable proposition. Besides, above-mentioned
calculations of the phonon spectra and electron-phonon coupling implicitly
account for the large susceptibility of the As$^{-4}$ ions (which comes
mostly from the outer, valence shell) yet they find no manifestation of
strong As polarization: neither particular phonon softening nor strong
coupling with any phonon.

\section{Pairing symmetry: general considerations}

\subsection{Geometrical consideration: excitation vectors and Fermi surface}

Given such disparate views that different researchers hold about the origin
of magnetism in ferropnictides and of the character of spin fluctuations
there, it may seem strange that a great majority of model calculations
predict the same pairing symmetry, $s_{\pm },$ with full gaps in both
electron and hole bands, but with the opposite signs of the order parameters
between the two. In fact, this is not surprising at all. To begin with, let
us point out that the sign of the interaction mediated by boson exchange is
always positive (attraction) for charge excitations (phonons, plasmons,
polarization excitons), since the components of a Cooper pair have the same
charge, but can be either positive (for triplet pairing, where the electrons
in the pair have the same spin) or negative (repulsion) for singlet pairing,
for spin excitations. That is to say, exchange of spin fluctuations mediates
repulsion. A quick glance at the anisotropic BCS equation reveals that
repulsive interactions can be pairing when, and only when the wave vector of
such a fluctuation spans parts of the Fermi surface(s) with opposite signs
of the order parameter (equivalently, one can say that an interaction that
is repulsive everywhere in the momentul space, can be partially attractive
in the real space, for instance, for electrons located an nearest lattice
sites).

\begin{figure}[tbp]
\includegraphics[width=0.9\linewidth
,viewport=0 350 700 600,clip
]{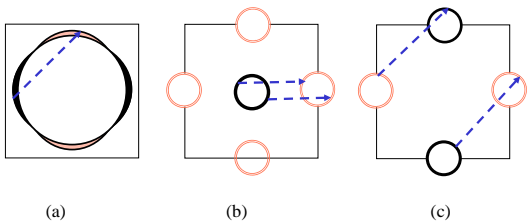}
\caption{(color online) (a) A cartoon illustrating how a repulsive
interaction corresponding to superexchange spin fluctuations $Q=(\protect\pi %
,\protect\pi $) may generate $d$-wave pairing in cuprates. (b) The same, for
an $s_{\pm }$ state and spin fluctuations with $Q=(\protect\pi ,0)$ (in a
Brillouin zone corresponding to one Fe per cell). (c) If the central hole
pocket is absent, the superexchange interaction favors a nodeiless $d$
state. }
\label{cupr}
\end{figure}

This can be illustrated on a popular model of high-$T_{c}$ cuprates, which
considers a simplified cylindrical Fermi surface nearly touching the edge of
the Brillouin zone and superexchange-driven spin fluctuations with the wave
vector $(\pi ,0)$. As Fig. \ref{cupr}a illustrates, such an interaction is
pairing in the $d_{x^{2}-y^{2}}$ symmetry, because it spans nearly perfectly
the lobes of the order parameter with the opposite signs. 

Most models used for ferropnictides assume a simplified fermiology with one
or more hole FSs and one or more electron FSs displaced by the SDW vector ($%
\pi ,0$) (in this Section, we use the notations corresponding to the
Brillouin zone with one Fe per cell). Any spin-fluctuation induced
interaction with this wave vector, no matter what the origin of these
fluctuations (FS nesting, frustrated superexchange, or anything else)
unavoidably leads to a superconducting state with the opposite signs of the
order parameter for the electrons and for the holes. Depending on the
details of the model the ground state maybe isotropic or anisotropic and the
gap magnitudes on the different sheets may be the same or may be different,
but the general extended $s$ symmetry with the sign-reversal of the order
parameter (an $s_{\pm }$ state) is predetermined by the fermiology and the
spin fluctuation wave vector (Fig. \ref{cupr}b).

It is worth noting that while most (but not all) models consider spin
fluctuations corresponding to the observed instability to be the leading
pairing agent, some include spin fluctuations of different nature [for
instance, nearest neighbor superexchange or nesting between the ``X'' and
``Y'' electron pockets, both corresponding to the same wave vector, ($%
\pi,\pi)$ in the unfolded zone and $(0,0)$ in the conventional zone], or
phonons, or direct Coulomb repulsion; these additional interactions may
modify the gap ratios and anisotropies (in extreme cases, creating nodes on
some surfaces), but, for a realistic choice of parameters, unlikely to
change the symmetry.

Moreover, if the radius of the largest FS pocket is larger than the magnetic
vector, spin fluctuations start to generate an \textit{intraband}
pair-breaking interaction, which by itself will lead to an angular
anisotropy and possible gap nodes.

The above reasoning, however, is heavily relying upon an assumption that the
topology predicted by the DFT is correct. So far, as discussed above, the
evidence from ARPES and from quantum oscillations has been favorable. It is
still of interest to imagine, for instance, electron-doped compounds not
having hole pockets at all or having them so small that the pairing energy
for them is negligible. It was pointed out\cite{Kuroki08,Gorkov} that in
this case spin fluctuations with different momentum vectors dominate and
create a nodeless $d$-wave state in the electron pockets, as Fig. \ref{cupr}%
c illustrates.

\subsection{General properties of the $s_{\pm}$ state}

Since the $s_{\pm }$ states constitute the most popular candidate for the
superconducting symmetry of pnictides, it is worth recapitulating the
physics of this state. Let us start with the simplest possible case: two
bands (two Fermi surfaces) and interband repulsive interaction between the
two. Let the interaction strength be $-V,$ and the DOSs $N_{1}\neq N_{2}.$
To be specific, let $N_{2}=\alpha N_{1},$ $\alpha \geq 1.$ Then in the weak
coupling limit the BCS equations read%
\begin{align}
\Delta _{1}& =-\int d\epsilon \frac{N_{2}V\Delta _{2}\tanh (E_{2}/2k_{B}T)}{%
2E_{2}}  \notag \\
\Delta _{2}& =-\int d\epsilon \frac{N_{1}V\Delta _{1}\tanh (E_{1}/2k_{B}T)}{%
2E_{1}}  \label{BCS}
\end{align}%
where $E_{i}$ is the usual quasiparticle energy in band $i$ given by $\sqrt{%
(\epsilon -\mu )^{2}+\Delta _{i}^{2}}.$ Near $T_{c}$ linearization gives%
\begin{align}
\Delta _{1}& =\Delta _{2}\lambda _{12}\log (1.136\omega _{c}/T_{c})  \notag
\\
\Delta _{2}& =\Delta _{1}\lambda _{21}\log (1.136\omega _{c}/T_{c}),
\label{BCSTc}
\end{align}%
where $\lambda _{12}=N_{2}V$, the dimensionless coupling constant, with a
similar expression for $\lambda _{21}.$ These equations readily yield $%
\lambda _{eff}=\sqrt{\lambda _{12}\lambda _{21}}$ and $-\Delta _{1}/\Delta
_{2}=\sqrt{N_{2}/N_{1}}\equiv \sqrt{\alpha }.$ Note that the Fermi surface
with the larger DOS has a smaller gap. It can also be shown that the gap
ratio at zero temperature in the weak coupling limit is also given by $\sqrt{%
N_{2}/N_{1}},$ and strong coupling effects tend to reduce the disparity
between the gaps.

The situation becomes more interesting for more than two orbitals with
distinct gaps. Let us consider a model for the hole-doped 122 compound. The
calculated FS (Fig.\ref{FS122}) shows three sets of sheets: Two e-pockets at
the corner of the zone, two outer h-pockets, formed by the $xz$ and $yz$
orbitals (degenerate at $\Gamma $ without the spin-orbit), and the inner
pocket formed by $x^{2}-y^{2}.$ In the DFT calculations all three hole
cylinders are accidentally close to each other, however, ARPES shows two
distinct sets, the inner barrel, one of which presumably corresponding to $%
x^{2}-y^{2}$ band, and the outer one, presumably $xz/yz.$ The pairing
interaction between the e-pockets and the two different types of the
h-pockets need not be the same (by virtue of the the matrix elements). Using
the same partial DOS as listed above for Ba$_{1.6}$K$_{0.6}$Fe$_{2}$As$_{2}$
(both total and individual DOS depend weakly on the position of the Fermi
level, reflecting the 2D character of the band structure at this doping),
roughly 1.2 st/eV for each hole band and the same for the two e-band
together, we get the coupling matrix 
\begin{equation}
\left( 
\begin{array}{ccc}
0 & 0 & -\lambda _{1}\nu _{1} \\ 
0 & 0 & -\lambda _{2}\nu _{2} \\ 
-\lambda _{1} & -\lambda _{2} & 0%
\end{array}%
\right) ,
\end{equation}%
where $\nu _{1,2}$ is the ratio of DOS of the first ($xz/yz)$ and the second
($x^{2}-y^{2})$ hole bands to that of the electron bands. Note that $\nu
_{1}\sim 2$ and $\nu _{2}\sim 1.$ Diagonalizing this matrix we find the gap
ratios to be $\Delta _{1}:\Delta _{2}:\Delta _{e}=\lambda _{1}:\lambda _{2}:%
\sqrt{\lambda _{1}^{2}\nu _{1}+\lambda _{2}^{2}\nu _{2}.}$ The latest ARPES
measurements\cite{NakayamaARPES} imply that $\Delta _{i}:\Delta _{o}\approx
2:1,$ where $i$ and $o$ stand for the inner and outer sets of hole Fermi
surfaces. This would mean that the two coupling constants are twice larger
that the other (although we do not know which), which is fairly possible.
However, that implies that the electron FS has a gap that is larger than
that of the largest hole band by at least a factor of $\sqrt{1.5}=1.22$
(assuming that the outer FSs  in the calculations, are formed by the $xz/yz$
bands; the opposite assumptions leads to an even larger electron-band gap).
This is in some disagreement with the ARPES data that suggest that $\Delta
_{e}$ is on the order of $\Delta _{i}$ or slightly smaller. However, this is
a small discrepancy, which can be easily corrected by introducing small
intraband electron-phonon coupling for the hole bands, and/or taking into
account possible gap suppression by impurities in the electron band. It is
also worth noting that the spread of the measured values, depending on the
sample and on the location on the FS, is on the order of 10\%.

\subsection{Coulomb avoidance}

It was realized quite some time ago that a $d$-wave pairing has an
additional advantage compared to an $s$-wave, namely that the electrons in a
Cooper pair avoid each other (the pair wave function has zero amplitude at $%
\mathbf{r-r}^{\prime }=0$), strongly reducing their local Coulomb repulsion.
The leading contribution to the pairing interaction in the single band
Hubbard model $U\sum_{\mathbf{k}}\left\langle c_{\mathbf{k\uparrow }}c_{-%
\mathbf{k\downarrow }}\right\rangle $ is repulsive, but vanishes as $\sum_{%
\mathbf{k}}\Delta _{\mathbf{k}}=0$ due to the symmetry of the $d$-wave
state. Thus, a contact Coulomb repulsion does not affect $d$-wave
superconductivity.

The simplest possible $s^{\pm }$-wave function is given by Eq.\ref{s+-}. In
this case, the sum over the Brillouin zone vanishes again due to nodes at $%
\pm ak_{x}\pm ak_{y}=\pi /2$. This description is however somewhat
misleading because it may produce a false impression that there is a
symmetry reason for the vanishing of the Coulomb repulsion in the $s^{\pm }$%
state, or that this particular functional form is essential for avoiding the
Coulomb repulsion. To illustrate that this is not the case, it is
instructive to consider a toy problem in reciprocal space. In the weak
coupling regime, the effective coupling matrix $\Lambda _{\mathbf{kk}%
^{\prime }}$ (note that the band index is uniquely defined by the wave
vector) is 
\begin{equation}
\Lambda _{\mathbf{kk}^{\prime }}=\lambda _{\mathbf{kk}^{\prime }}-\mu _{%
\mathbf{kk}^{\prime }}^{\ast },
\end{equation}%
where $\lambda $ is the original coupling matrix in orbital space and $\mu _{%
\mathbf{kk}^{\prime }}^{\ast }$ is the renormalized Coulomb pseudopotential.
The critical temperature is determined by the largest eigenvalue of the
matrix $\Lambda ,$ and the $\mathbf{k}$ dependence of the order parameter $%
\Delta _{\mathbf{k}}$ is given by the corresponding eigenvector. If $\mu
^{\ast }$ is a constant and $\sum_{\mathbf{k}}\Delta _{\mathbf{k}}=0$ (as in
the $d$-wave case), any eigenvector of the matrix $\lambda $ is also an
eigenvector of $\Lambda ,$ with the same eigenvalue. This proves that
Coulomb avoidance takes place for any superconductor where the order
parameter averages to zero over the entire FS, and not only for the $d$-wave
symmetry.

Let us now consider a specific $s^{\pm }$ superconductor. For simplicity,
let us take two bands with the same DOS, $N_{1}=N_{2}=N$ and with an
interband coupling only:%
\begin{equation}
\lambda _{ij}=\left( 
\begin{array}{cc}
0 & -VN \\ 
-VN & 0%
\end{array}%
\right) .
\end{equation}%
We shall also assume that the Coulomb repulsion $U$ is a contact
interaction, so that $\mu _{ij}^{\ast }=UN$ is the same for all matrix
elements. The maximal eigenvalue of $\Lambda $, which corresponds to the
effective coupling constant $\lambda _{\mathrm{eff}}$, is indeed simply $VN$
and\emph{\ independent} of $U$. The corresponding eigenvector is $\Delta
_{1}=-\Delta _{2}$, i.e. the $s^{\pm }$ state. The Coulomb interaction is
irrelevant, just like in case of $d$-wave pairing. The effect is however a
consequence of the assumed symmetry of the two bands. In general, unlike the
d-wave, no symmetry requires that $\sum_{\mathbf{k}}\Delta _{\mathbf{k}}=0$.
This can already be seen if one considers a model with distinct densities of
states: $N_{2}=\alpha N_{1}=\alpha N$. We have 
\begin{equation}
\lambda _{ij}=\left( 
\begin{array}{cc}
0 & -\alpha VN \\ 
-VN & 0%
\end{array}%
\right) .
\end{equation}%
and the weak-coupling gap ratio near $T_{c}$ is $\sqrt{\alpha }$. Now the
effect of the Coulomb repulsion is not nullified, but is still strongly
suppressed. The eigenvalues are easily determined. The key result is that
the maximal eigenvalue remains positive for all finite $\alpha $. Even the
extreme limit $\lambda _{\mathrm{eff}}^{\pm }(U\rightarrow \infty
)=2VN\alpha /(1+\alpha )$ is for realistic $\alpha $ only somewhat reduced
compared to $\lambda _{\mathrm{eff}}^{\pm }(U=0)=\sqrt{\alpha }VN$. This is
qualitatively different from the regular ($s_{++})$ interband-only pairing
with an attractive interband interaction of the same strength. In this case, 
$\lambda _{\mathrm{eff}}^{++}(U>V/2)<0$, and the Coulomb interaction
dominates over the attractive interband pairing interaction. In the linear
in $UN$ regime, the suppression rate of $\lambda _{eff}(U)$ is $(\sqrt{%
\alpha }-1)/2$ for $s^{\pm }$ and $(\sqrt{\alpha }+1)/2$ for $s^{++}$
pairing. For example, for the DOSs ratio of $4$ (the gap ratio is then $2$) $%
\mu ^{\ast }\approx 0.25\lambda _{eff}\left( U=0\right) $ will suppress an $%
s^{++}$ superconductivity entirely, while in the $s^{\pm }$ case the
effective coupling will be reduced only by 8\%.

The efficiency of the Coulomb avoidance is neither limited to the assumption
of a uniform Coulomb interaction among and within the bands, nor is a result
of the weak coupling approach. Strong coupling FLEX type calculations also
find pairing states with very small repulsive contribution due to Coulomb
interaction\cite{Yao,Sknepnek}.

\section{Pairing symmetry: experimental manifestations}

\subsection{Parity}

Since we want to review the experimental situation regarding the pairing
symmetry, the first question to ask is, whether superconductivity is singlet
or triplet? Fortunately, this question can be answered relatively
confidently. Measurements of the Knight shift on single crystals of the
Co-doped BaFe$_{2}$As$_{2}$ superconductor\cite{NMR-Co} clearly indicate
full suppression of spin susceptibility in the superconducting state \textit{%
in all directions}, incompatible with a triplet pairing in a tetragonal
crystal. For other compounds only polycrystalline, direction-averaged data
exist, but they fully agree with the above result, virtually excluding
triplet superconductivity. This leaves, of all possible scenarios,
essentially three: conventional $s$ (presumably multigap), $s_{\pm }$ and $d$%
.

\subsection{Gap amplitude}

All experiments that distinguish between different pairing states can be,
roughly speaking, grouped into two classes: those probing the gap amplitude
and those probing the gap symmetry. The advantage of the former is that they
are comparatively easier to perform. The temperature dependence of any
observable sensitive to the excitation gap is sensitive to the presence of
nodes or multiple gaps. The disadvantage is that only a measurement of the
relative phase of the wave function will unambiguously determine the pairing
state, including its symmetry.

Important and very transparent probes of the gap amplitude are thermodynamic
measurements. The early reports of the specific heat leaned towards
power-law behavior characteristic of nodal superconductivity. The latest
data \cite{SH1,SH2} suggest a fully gapped superconductivity, or a dominant
fully gapped component with possible small admixture of a nodal state. While
the experimental situation is still far from consensus, especially regarding
the 1111 family, a few observations may be in place: (i) The specific heat
jump in the h-doped BaFe$_{2}$As$_{2}$ is strong and sharp, and in 1111
compounds is weak and poorly expressed. This cannot be ascribed to a
difference in calculated band structures. This is either due to sample
quality issues or possibly to the more isotropic character of
superconducting and magnetic properties in 122 systems. (ii) In no case can
specific heat temperature dependence be fitted with one gap. Multiple gap
fits, having more parameters, are of course less reliable. (iii) Another,
usually more reliable signature of nodal superconductivity is a square-root
dependence of the specific heat coefficient on the magnetic field. Existing
reports\cite{SH1} however show a clear linear dependence, characteristic of
a fully gapped superconductor.

Another popular probe is temperature dependence of the NMR relaxation rate.
Extensive studies have been done in this aspect (see other articles in this
volume). In all studied systems, the relaxation rate is non-exponential. The
initial impression was that the relaxation rate is cubic in temperature, $%
1/T_{1}\propto T^{3},$ consistent with nodal lines\cite{Terasaki,Ishida}.
Later it was argued that the data cannot be described by a single power law
as in the cuprates\cite{Zheng,Kobayashi}. These results were obtained for
the 1111 systems. The situation with the 122 family is even less clear.
Published data\cite{Fukazawa,NMR-Co} do not show exponential decay either,
but the results are equally far from any single power law behavior. Even
more puzzling, the only paper reporting on the low-$T_{c}$ LaFePO
superconductor claims that the relaxation rate does not decrease below $T_{c}
$ at all\cite{NMR-P}.

The third relevant experiment is measuring the London penetration depth.
Reports are again contradictory. For instance, in Pr-based 1111 compound the
penetration depth was found\cite{Hashimoto} to barely change between $%
\approx 0.05T_{c}$ and $T^{\ast }\approx 0.35T_{c},$ and than increase
roughly as $(T-T^{\ast })^{2}$ between $T^{\ast }$ and $\approx 0.65T_{c},$
a picture roughly consistent with a multi-gap nodeless superconductor.
Malone $et$ $al$\cite{Tony} measured Sm-based 1111 and were able to fit
their data very well in the entire interval from $T_{c}/30$ and $T_{c}$
using two full gaps. In Nd-based 1111 the penetration depth was measured at $%
T>0.1T_{c}$ and fitted with a single anisotropic gap for $0.1T_{c}<T<T_{c}/3$%
,\cite{MartinNd} however, the latest result from the same authors, taken at
lower temperature, can be better fitted with a quadratic law\cite{Ruslan}.
Similar quadratic behavior has been clearly seen in the 122 compounds\cite%
{Ruslan122}. At the same time, the low-$T_{c}$ LaFePO is again odd: it shows
a linear behavior\cite{Fletcher}.

To summarize, the thermodynamic data on average lean towards a nodal
superconductivity. However, \textit{some} data are not consistent with the
gap nodes, and there is no clear correlation with the sample quality either
way. Moreover, while some data suggest line nodes, others are consistent
only with point nodes, in the clean limit. One can say with a reasonable
degree of confidence that the entire corpus of the data cannot be described
by any one scenario in the clean limit. On the other hand, essentially any
temperature dependence of thermodynamic characteristics can be fitted if a
particular distribution of impurity scattering is assumed in an intermediate
regime between the Born and the unitary scattering, and a particular
relation between the intra- and interband scattering (there have been a
number of paper doing exactly that for the NMR relaxation rate, for
instance, Ref. \cite{Parker}, or for the penetration depth, for
instance, Ref. \cite{ChubukovPD}). However, the fact that all these papers
rely upon specific combinations of parameters, while the phenomena they seek
to describe are rather universal, calls for caution. Besides, except in the
pure unitary regime, scattering is accompanied by a $T_{c}$ suppression and
most papers do not find any correlation between thermodynamic probes and $%
T_{c}$ among different samples. Another possibility is that required
scattering is provided not by impurities, but by intrinsic defects that are
thermodynamically or kinetically necessarily present in all samples (for
example, dynamic domain walls introduced in Ref. \cite{domains}). More
measurements at the lower temperature and on clean samples will probably
clarify the matter. At the moment one cannot consider this problem solved.

Close to the thermodynamic measurements are tunneling type experiments. As
of now, these have been nearly exclusively point-contact Andreev reflection
probes. Here, again, the experimental reports are quite inconsistent,
moreover, the situation is in some sense worse than in thermodynamic probes,
since uncontrollable surface properties enter the picture. Interpretation
generally includes fitting one curve with a large number of parameters, and
the procedure is not always well defined. Generally speaking, three types of
results have been reported: $d$-wave like, single full gap-like, and
multigap. Interpretation is particularly difficult because within the $%
s^{\pm }$ picture formation of subgap Andreev bound states was predicted (%
\textit{e.g., }Refs. \cite{Brinkman,Hu-Jos}) that can be easily mistaken for
multiple gaps.

\subsection{Phase-sensitive probes}

In view of all that, experiments directly probing the gap symmetry are
highly desirable. The paramagnetic Meissner effect, also known as \textit{%
Wohlleben effect}, occurs in a polycrystalline sample when inter-grain weak
links have random order parameter phase shifts, $0$ or $\pi .$ It has been
routinely observed in cuprates and is considered a key signature of $d$-wave
superconductivity. The effect does not exist in conventional, even
anisotropic and multi-gap superconductors, even though sometimes it can be
emulated by impurity effects in the junctions. For $d-$wave superconductors
without pronounced crystallographic texture the Wohlleben effect is
expected, and its absence can be taken as evidence against $d$-wave.
Finally, in the $s^{\pm }$ scenario the phase is the same by symmetry for $%
(100)$ and $(010)$ grain boundaries, and there are good reasons to expect
the same phase for $(110)$ boundaries as well. There may or may not be a $%
\pi $ phase shift for phase boundaries at some specific orientation, likely
for a narrow range of angles\cite{parker-phases}, but probably not enough to
produce a measurable Wohlleben effect. The absence of the effect in
experiment\cite{KAM} is a significant argument against $d$-wave, but hardly
helps to distinguish $s$ from $s^{\pm }.$

Similarly, the $c$-axis tunneling provides evidence against the $d$-wave,
where the Josephson current strictly parallel to the crystallographic $c$
direction vanishes by symmetry. Experimentally a sizable current was found%
\cite{RickGreene}.

Recalling the cuprates again, the ultimate argument in favor of the $d$-wave
was provided by the corner Josephson junction experiments that probe
directly the phase shift between two separate junctions; in cuprates, with
their $d_{x^{2}-y^{2}}$ symmetry, these junction were to be along the $(100)$
and $(010)$ directions. Similarly, a potential $d_{xy}$ state could be
detected by the combination of $(110)$ and $(\bar{1}10)$ directions. On the
other hand, \ a conventional $s$ state would not produce a phase shift for
any combination of contacts. Again, the case of $s_{\pm }$ superconductivity
is nontrivial. While symmetry does not mandate a $\pi $ shift for any
direction, it can be shown that, depending on the electronic structure
parameters and properties of the interface, there may exist intermediate
angles (between $0$ and $45^{o})$ where a $\pi $ shift is possible\cite%
{parker-phases}. It also may be possible if the two junctions have different
tunneling properties, so that one of them filters through only hole-pocket
electrons, and the other only electron-pockets. It is not as bizarre as it
may seem, and some possibilities were discussed in Ref. \cite{parker-phases}%
. Probably the most promising design involves \textquotedblleft
sandwiches\textquotedblright\ of various geometries. The first proposal of
that kind was by Tsoi et al\cite{Hu-Jos}, who suggested an $s/s^{\pm
}/s^{\prime }$ trilayer, where $s$ is a conventional quai-2D superconductor
with a large Fermi surface that has no overlap with the hole FS of the $%
s^{\pm }$ layer (equivalently, a superconductor with small Fermi surfaces
centered around the M points), and $s^{\prime }$ is a conventional
superconductor with a small FS centered around $\Gamma .$ This was followed
by another proposal of a bilayer of  hole-doped and electron-doped 122
materials\cite{parker-phases}. In both cases the idea is that the current
through the top of the sandwich will be dominated by the electron FS, and
through the bottom by the hole one. Both proposals require momentum
conservation in the interfacial plane, that is, basically, epitaxial or very
high quality interface. The former proposal has an additional disadvantage
of requiring \textit{two} high-quality interfaces with very special
conventional superconductors, particularly the one that should filter
through the electron FS is rather difficult to find. As of now, no
experiments have been reported pursuing any of the above suggestions, but
with better single crystals and thin films it should become increasingly
doable. It should be stressed, however, that in this case, unlike the
cuprates, an absence of the $\pi $ shifts in any of the proposed geometries
does not disprove the $s^{\pm }$ scenario, since the effect here is
quantitative rather than qualitative, but the presence of the sought effect
would be a very strong argument in favor of it. On the other hand, standard
90$^{o}$ corner junction experiments similar to cuprates are also important,
as they could prove unambiguously that the symmetry is not $d$-wave (even
though they cannot distinguish between $s$ and $s^{\pm }).$

Further properties of interfaces between an $s^{\pm }$ superconductor and
normal metal or conventional superconductor are now actively being studied
theoretically, encouraging further experimental research. Probably we will
see first results within the next year.

\subsection{Coherence factor effects}

Other signatures of the $s^{\pm }$ state are based on the fact, previously
pointed out by many in connection with the cuprates, that the coherence
factors are \textquotedblleft reversed\textquotedblright\ for electronic
transitions involving order parameters of the opposite sign. In the
conventional BCS theory, as is well known, coherence factors of two kinds
appear. The first kind, sometimes called \textquotedblleft Type
I\textquotedblright\ or \textquotedblleft minus\textquotedblright\ coherence
factor, is given by the expression $(1-\Delta _{\mathbf{k}}\Delta _{\mathbf{k%
}^{\prime }}/E_{\mathbf{k}}E_{\mathbf{k}^{\prime }}),$ where $E_{\mathbf{k}}=%
\sqrt{\Delta _{\mathbf{k}}^{2}+\varepsilon _{\mathbf{k}}^{2}},$ and $%
\varepsilon _{\mathbf{k}}$ in the normal state excitation. The other kind,
Type II or the \textquotedblleft plus\textquotedblright\ coherence factor
has the opposite sign in front of the fraction. If both order parameters
entering this formula have the same sign, the Type I factor is destructive,
in the sense that it goes to zero when $\varepsilon \rightarrow 0,$ and
cancels out the peak in the superconducting DOS. Type I factors appear, for
instance, in the polarization operator, and as a result there are no
coherence peaks in phonon renormalization (as measured by ultrasound
attenuation, for instance) and in spin susceptibility (including the Knight
shift). Type II factors appear, for instance, in the NMR relaxation rate,
and they are constructive, resulting in the famous Hebel-Slichter peak below 
$T_{c}.$

Obviously, if $\Delta _{\mathbf{k}}$ and $\Delta _{\mathbf{k}^{\prime }}$
have opposite signs, the meaning of the coherence factors is reversed; the
Type I factors are now constructive and the Type II destructive. There are
several straightforward ramifications of that. For instance, as it was
pointed out already in the first paper proposing the $s^{\pm }$ scenario\cite%
{mazin08}, the spin susceptibility at the SDW wave vector should show
resonance enhancement just below $T_{c}$. For explicit calculations of this
effect see for example Refs.\cite{Maier08,Ereminres}. There are indeed some
reports of this effect, as measured by neutron scattering\cite{Christ}. In
principle, one can expect a similar effect in the phonon line-width, for the
phonons with the same wave vector, just below $T_{c},$ but this is really
hard to observe.

Less straightforward are cases of the quantities that involve averaging over
the entire Brillouin zone, in which case the answer, essentially, depends on
which processes play a more dominant role in the measured quantity, those
involving intra-, or interband scattering. The answer usually depends on
additional assumptions about the matrix elements involved, which can rarely
be calculated easily from first principles. An example is electronic Raman
scattering; a possibility of a resonant enhancement in some symmetries has
been discussed recently\cite{Raman}.

\section{Role of impurities}

Impurity and defect scattering is believed to play an important role in
pnictide superconductors. Proximity to a magnetic instability implies that
ordinary defects may induce static magnetic moments on the neighboring Fe
sites and thus trigger magnetic scattering. If, as is nearly universally
believed, an order parameter with both signs is present, nonmagnetic
impurities are also pair-breaking. Thus the anticipation is that in regular
samples, and maybe in samples of much higher quality, impurity-induced pair
breaking will play a role.

Our intuition regarding the impurity effects in superconductors is largely
based upon the Abrikosov-Gorkov theory of Born-scattering impurities in BCS
superconductors. There was an observation at that time that folklore
ascribes to Mark Azbel: Soviet theorists do what can be done as good as it
should be done, and American ones do what shall be done as good as it could
be done. For many years the approach to the impurity effects in
superconductors was largely Soviet: most researchers refine the
Abrikosov-Gorkov theory, applying it to anisotropic gaps and  to
unconventional superconductors, and relatively little has been done beyond
the Born limit --- despite multiple indications that most interesting
superconductors, from cuprates to MgB$_{2}$ to pnictides are in the unitary
limit or in an intermediate regime.

The physics of the nonmagnetic scattering in the two different limits is
quite different. In the Born limit, averaging over all scattering events
yields a spatially uniform superconducting state and tries to reduce the
variation of the order parameter over the FS. Ultimately, for sufficiently
strong scattering, the order parameter becomes a constant, corresponding to
the DOS-weighted average over the FS. Note that unless this average is zero
by symmetry (like in d-wave) the suppression of $T_{c},$ while linear at
small concentrations, is never complete. As pointed out by Mishra $et$ $al.$ 
\cite{Scalapino2009}, this effect should manifest itself most clearly in an
extended s-wave pairing with accidental nodes in the order parameter.
Indeed, while in $d$-wave superconductors impurities broadens nodes into
finite gapless spots, in an extended $s$ case it is likely that the order
parameter of one particular sign dominates a given FS pocket, in which case
Born impurities will first make the parts of the FS with the
\textquotedblleft wrong\textquotedblright\ order parameter gapless, and then
lead to a fully gapped superconductivity. Of course, this only holds for
nonmagnetic impurities. Isotropic magnetic impurities will be just
pair-breaking as they are in conventional superconductors, with the only
interesting new physics being that magnetic impurities cease being
pair-breakers if they scatter a pair such that the sign of the order
parameter is flipping. The rule of thumb is that a scattering path for which
magnetic scattering is pair-breaking (no change of sign of the order
parameter), nonmagnetic scattering will not be pair-breaking, and $vice$ $%
versa.$

The physics of the unitary limit is quite different. In that limit, the
concentration of impurities is relatively low, but the scattering potential
of an individual impurity is strong, $N(0)v_{imp}\gg 1.$ In that case rather
than suppressing superconductivity uniformly each impurity creates a bound
state at the chemical potential, thus creating a zero energy peak in the
density of states, without substantial suppression of the bulk
superconductivity. Increasing the impurity concentration broadens the peak,
while increasing its strength barely has any effect at all \cite{Muzikar}.
In an intermediate case between the Born limit and the unitary limit, the
bound state is formed inside the gap at a finite energy and is the broader
the closer it is to the gap (that is, closer to the Born limit).

The principal difference from the point of view of the experiment is that
the unitary or intermediate scattering can create subgap density of states
at arbitrary low energy at any temperature, without a drastic suppression of 
$T_{c}.$ It was shown in Ref. \cite{Parker} that any standard code for
solving the Eliashberg equations in the Born limit can be easily modified,
with minor changes, to treat the unitary limit, as well as any intermediate
regime. Therefore we anticipate an imminent shift in the community from the
\textquotedblleft Soviet\textquotedblright\ approach to the
\textquotedblleft Western\textquotedblright\ approach, with more
quantitative understanding of the effect beyond the Born approximation.

\section{Conclusions}

In this article we presented a brief overview of some proposals that have
been made for the pairing state in the Fe-pnictide superconductors. In
particular, we summarized arguments that support the view that the vicinity
of superconductivity and magnetism in these systems is not accidental. The
obvious appeal of this, and essentially any other electronic pairing
mechanism is, of course, that the involved energy scales, and thus $T_{c}$,
can in principle be larger if compared to pairing due to electron-phonon
interaction. Electronic mechanisms also promise a new level of versatility
in the design of new superconductors.

At this early stage in the research on the iron pnictide family, experiments
have not conclusively determined the pairing symmetry, the detailed pairing
state or the microscopic pairing mechanism. Still, in our view a plausible
picture emerges where superconductivity is caused by magnetic fluctuations.
Only two ingredients are vital to arrive at a rather robust conclusion for
the pairing state. First, pnictides need to have Fermi surface sheets of two
kinds, one near the center of the Brillouin zone, and the other near the
corner. Second, the typical momentum for the magnetic fluctuations should be
close to the ordering vectors $\mathbf{Q=}\left( \pi,\pi\right) $ of the
parent compounds. Then, magnetic interactions lead quite naturally to an
efficient inter-band coupling that yields an $s^{\pm}$ pairing state. This
result is general in the sense that it is obtained regardless of whether one
develops a theory based on localized quantum magnetism or itinerant
paramagnons. There is evidence that the two needed ingredients are present
in the pnictides. Fermi surface sheets at the appropriate locations have
been predicted in non-magnetic LDA calculations and seen in ARPES
experiments. The magnetic ordering vector has been determined via neutron
scattering, even though we have to stress that a clear identification of
magnetic fluctuations for superconducting systems without long range
magnetic order is still lacking.

The resulting $s^{\pm }$ pairing state has a number of interesting
properties. As far as the a group theoretic classification is concerned, its
symmetry is the same as that for a conventional $s$-wave pairing, where the
gap-function has same sign on all sheets of the Fermi surface. However,
there are significant differences between the two states. The sign change in
the gap affects the coherence factors, leading to the resonance peak in the
dynamic spin susceptibility and the absence of a Hebel-Slichter peak in NMR.
Nonmagnetic impurities affect the $s^{\pm }$-state just like magnetic
impurities do in an ordinary $s$-wave state, i.e. here a behavior more akin
to $d$-wave superconductors. Another implication of the sign change in the $%
s^{\pm }$-state leads to rather efficient Coulomb avoidance.

The presence of nodes in the superconducting gap in still an open issue. In $%
d$-wave or $p$-wave pairing states, nodal lines or points are fixed by
symmetry. This is different for the $s^{\pm }$-state. In its most elementary
version, the sign change of the gap corresponds to a node located between
two Fermi surface sheets. This is the case for the $\Delta \left( \mathbf{k}%
\right) $ given in Eq.\ref{s+-}. Energetic arguments favor such a gapless
state as long as the momentum transfer $\mathbf{Q}$ couples efficiently to
large parts of distinct Fermi surface sheets and Coulomb avoidance is
efficient. However, as there is no symmetry constraint for the location of
the nodes, it is in principle possible that there are nodes on some Fermi
surface sheets.

Next to the nature of the pairing state,  the microscopic understanding of
the magnetism of the Fe-pnictides is one of the most interesting aspects of
these materials. Are these systems made up of localized spins that interact
via short ranged, nearest neighbor exchange interactions or, are they better
described in terms of itinerant magnetism? While we emphasized that many
aspects of the pairing state emerge regardless of which of these points of
view is correct, this is really only true for the most elementary aspects of
the theory. As our understanding of these materials deepens, dynamical
aspects of the pairing state will become more and more important, and the
details of the magnetic degrees of freedom will matter. In our view, the
most sensible description starts from itinerant electrons, however with
significant electron-electron interaction. In detail, we find numerous
arguments that emphasize the role of magneto-elastic couplings and that
favor a sizable Hund coupling, i.e. the multi orbital character and the
corresponding local multi-orbital interactions are important to understand
the magnetism and superconductivity alike. Regardless of whether this
specific point of view is correct or not, it is already evident that the
ferropnictides make up a whole new class of materials that stubbornly refuse
to behave according to one of the simple minded categories of condensed
matter theory.

\section{Acknowledgements}

This research was supported by the Ames
Laboratory, operated for the U.S. Department of Energy by Iowa State
University under Contract No. DE-AC02-07CH11358 (J.S.), and by the Office of Naval 
Research (I.I.M.). The authors wish to thank all their friends and collaborators,
without whom this works could not be accomplished, and their numerous 
colleagues who read the manuscript and sent us many useful and insightful comments.


\begin{thebibliography}{999}
\bibitem{Ref1} All original calculations used in this article were performed
using the standard full-potential LAPW package WIEN2k. The details can be
found in Refs. \cite{mazin08,Mazin08a}. Unless stated otherwise, the
experimental crystal structure is used.
\bibitem{Yares} Not only the energetics of collinear LDA calculations cannot
be mapped upon a $J_1+J_2$ model\cite{Mazin08a} (which required forced FM
calculations, it was also directly demonstrated by noncollinear LDA
calculations by Stuttgart group\cite{Yaresko}.
\bibitem{Ref2} DFT calculations\cite{Ref1} yield the following magnetic
stabilization energies for FeTe: 332 (227) meV/Fe with (without) spin-orbit
coupling in the experimental P21/m double-stripe structure, and 281 (205)
meV/Fe with in the single-stripe Pccm structure (with full optimization of
atomic positions in each structure). Thus, the experimental structure
appears to be below the competing structure by 22 (or 49, with spin- orbit)
meV/Fe. Exactly the same calculation for LaFeAsO\cite{Mazin08a} render the
double-stripe structure well above the single-stripe one. The difference can
be traced down to the fact that for LaFeAsO only the single stripe structure
opens a pseudogap in DOS at the Fermi level, while for FeTe both AFM
structures open comparable pseudogaps. Further details will be published
elsewhere\cite{tbp}.
\end{thebibliography}
\end{document}